\documentclass[useAMS,usenatbib,twocolumn,floatfix]{mn2e}
\usepackage{natbib}
\usepackage{graphicx}
\usepackage{times}
\usepackage{rotate}
\usepackage{color,url}
\usepackage[
bookmarkstype=toc=true,
 linktocpage=true,
 bookmarks=true,
 unicode
 ]{hyperref}
\def\bs#1{\mbox{\boldmath $#1$}}

\newcommand{\simgt}{\lower.5ex\hbox{$\; \buildrel > \over \sim \;$}}
\newcommand{\simlt}{\lower.5ex\hbox{$\; \buildrel < \over \sim \;$}}

\newcommand{\bvec}[1]{\mbox{\boldmath $#1$}}

\newcommand{\beq}{\begin{equation}}
\newcommand{\beqa}{\begin{eqnarray}}
\newcommand{\eeq}{\end{equation}}
\newcommand{\eeqa}{\end{eqnarray}}

\newif\iffigure
\figuretrue

\begin{document}

\title[HSC Cluster $\gamma$-ray cross correlation]%
{Measurement of redshift dependent cross-correlation
  of HSC clusters and \emph{Fermi} $\gamma$ rays}

\author[Hashimoto et al.]
{Daiki Hashimoto $^1$,
Atsushi  J. Nishizawa$^{1,2}$,
Masato Shirasaki$^3$, 
Oscar Macias$^4$,
\newauthor
Shunsaku Horiuchi$^4$
Hiroyuki Tashiro$^1$, 
Masamune Oguri$^{5,6,7}$\\
$^{1}$ Department of Physics, Nagoya University, Nagoya, Aichi
464-8602, Japan\\
$^{2}$ Institute for Advanced Research, Nagoya University, Nagoya, Aichi
464-8602, Japan \\
$^{3}$ National Astronomical Observatory of Japan (NAOJ), 
Mitaka, Tokyo 181-8588, Japan \\
$^{4}$ Center for Neutrino Physics, Department of Physics, Virginia Tech, Blacksburg, Virginia
24061, USA\\
$^{5}$ Research Center for the Early Universe, University of Tokyo, Tokyo 113-0033, Japan\\
$^{6}$ Department of Physics, University of Tokyo, Tokyo 113-0033, Japan\\
$^{7}$ Kavli Institute for the Physics and Mathematics of the Universe (Kavli IPMU, WPI), University of Tokyo, Chiba 277-8582, Japan\\
}

\date{\today}

\maketitle

\begin{abstract}
The cross-correlation study of the unresolved $\gamma$-ray background~(UGRB)
with galaxy clusters has a potential to reveal the nature of the UGRB.
In this paper, we perform a cross-correlation analysis between
$\gamma$-ray data by the Fermi Large Area Telescope~(\emph{Fermi}-LAT) and
a galaxy cluster catalogue from the Subaru Hyper Suprime-Cam~(HSC) survey. 
The Subaru HSC cluster catalogue provides a wide and
homogeneous large-scale structure distribution out to the high redshift 
at $z=1.1$, which has not been accessible in previous cross-correlation studies.
We conduct the cross-correlation analysis not only for clusters in the
all redshift range~($0.1 < z < 1.1$) of the survey,
but also for subsamples of clusters divided into redshift bins, the low 
redshift bin~($0.1 < z < 0.6$) and the high redshift bin~($0.6 < z < 1.1$), 
to utilize the wide redshift coverage of the cluster catalogue. 
We find the evidence of the cross-correlation signals with the significance of 
2.0-2.3$\sigma$ for all redshift and low-redshift cluster samples.
On the other hand, for high-redshift clusters, we find the signal with 
weaker significance level~(1.6-1.9$\sigma$).
We also compare the observed cross-correlation functions with predictions of 
a theoretical model in which the UGRB originates from $\gamma$-ray emitters 
such as blazars, star-forming galaxies and radio galaxies.
We find that the detected signal is consistent with the model prediction.

~

\noindent {\bf Key~words:} cosmology: large-scale structure of Universe - observations: galaxy clusters - gamma-rays: diffuse background. 
\end{abstract}

\section{Introduction}
\label{sec:intro}
An extragalactic $\gamma$-ray background (EGB) is among 
the most important subjects in high-energy astrophysics. 
The EGB has been commonly estimated with subtraction of 
the diffuse Galactic $\gamma$ rays from the observed emission in 
various $\gamma$-ray observational programs \citep[e.g.][]{Thompson:2008, Atwood+:2009}.
The latest EGB measurement has been performed 
by Fermi Large Area Telescope~(\emph{Fermi}-LAT) 
in the whole sky except low galactic latitudes,
at the energy range from 100 MeV to 800 GeV \citep{Ackermann+:2015}.
At the same time, \emph{Fermi}-LAT has discovered $\sim 3000$ point sources \citep{Acero+:2015} and clarified the contribution of those sources to the EGB.
\citet{Ackermann+:2015} estimated the resolved point source contributions 
to the EGB emission to be $\sim$35 per~cent \citep[also see][]{2015ApJ...800L..27A}.
The remaining fraction in the EGB is referred to as an unresolved $\gamma$-ray background (UGRB), whose origin is still under debate. 

The UGRB is expected to be the sum of $\gamma$ rays
originating from various unresolved $\gamma$-ray sources and 
some diffuse processes \citep[see][for review]{2015PhR...598....1F}.
\citet{2015ApJ...800L..27A} showed that the mean intensity of the UGRB can be explained 
by cumulative $\gamma$-ray emissions from blazars, 
star-forming galaxies and radio galaxies.
To estimate the exact amount of $\gamma$-ray emissions 
from those unresolved astronomical sources,
one needs a precise modeling of $\gamma$-ray luminosity function 
and energy spectrum for each population,
which is still being developed \citep{Lacki+:2014, Hooper+:2016, Di_Mauro:2018}.
In addition to unresolved known $\gamma$-ray sources, 
some purely diffuse processes can account for the EGB emission. 
These diffuse processes include annihilation or decay of cosmic dark 
matter \citep[e.g.][]{1996PhR...267..195J},
evaporation of primordial black holes formed in the early Universe \citep[e.g.][]{2010PhRvD..81j4019C},
and shock radiation from the medium in galaxy clusters \citep[e.g.][]{2000Natur.405..156L}.

To study the origin of the UGRB, the measurement of mean intensity has been used so far,
whereas there is other information beyond the sky average in observed UGRB intensity.
The fluctuation in the UGRB is a powerful probe to study the statistical relation 
between the UGRB and large-scale structures in the Universe.
Two-point correlation of the UGRB has been measured in actual data set \citep[e.g.][]{Xia+2011, 2016PhRvD..94l3005F}, which can bring meaningful information about dark matter annihilation \citep[e.g.][]{2006PhRvD..73b3521A, 2013PhRvD..87l3539A, 2016PhRvD..94l3005F} 
and the properties of faint astrophysical sources \citep[e.g.][]{2017PhRvD..95l3006A}.
Furthermore, cross correlation measurements with various tracers of large-scale structures 
in the Universe have been performed \citep[e.g.][]{2014PhRvD..90f3502S, 2015ApJS..217...15X, 2015PhRvD..92l3540S, 2015ApJ...802L...1F, 2017ApJ...836..127F, Branchini+:2017}. 
Such cross correlations are naturally expected from 
the theory of standard structure formation in modern cosmology
since astronomical objects or particles inducing diffuse $\gamma$ rays 
are preferentially located in high density regions in the Universe.
Detailed measurement of cross correlation with various tracers 
of large-scale structures helps separating the contributions from 
different $\gamma$-ray emitters in the UGRB \citep[e.g.][]{2014JCAP...10..061A, 2015JCAP...06..029C}. 

Among various tracers of large-scale structures,
galaxy clusters are one of the most interesting probes because they are largest gravitationally-bound objects in the Universe and rich $\gamma$ rays may be confined in galaxy clusters \citep[e.g.][]{2000ApJ...545..572T, 2010MNRAS.409..449P}.　
Although $\gamma$-ray emission from galaxy clusters has not yet been detected on individual basis \citep[e.g.][]{2016ApJ...819..149A}, there exists {\it statistical} evidence supporting the idea that the spatial distribution of galaxy clusters correlates with observed UGRB.

\citet{Branchini+:2017} have studied the cross-correlation signal with the UGRB and three different galaxy cluster catalogues~ (WHL12, redMaPPer and PlanckSZ;~\citealt{Rykoff+:2014}, \citealt{Wen-Han:2015} and \citealt{Planck_Collaboration+:2016}). They have detected the cross correlation signals with $\sim6-7\sigma$ significance level for WHL12 and redMaPPer, and argued that compact sources like active galactic nuclei and star-forming galaxies have large contribution to the signal.

Moreover, larger galaxies or galaxy clusters catalogues that extend the redshift ranges 
to higher-redshifts have recently been constructed.
Thanks to their large number densities, one can perform the cross correlation 
measurement of the UGRB with subsample of galaxies or galaxy clusters divided 
by their redshifts. 
The information of the redshift dependence of the UGRB is expected to provide 
a new clue to the origin and nature of the UGRB.
For instance, \citet{Cuoco+:2017} measured cross-correlation signals 
tomographically using various galaxy catalogs with $0.05 < z < 3$.
They presented the signals with high significance in a wide redshift range.
These results demonstrate that tomographic approach indeed works with 
actual data set and is helpful to explore new physics such as dark matter 
annihilation or decays in the UGRB. 

In this work, we use a cluster catalogue for 
a wide redshift range of $0.1 < z < 1.1$ from the Subaru Hyper Suprime-Cam
(HSC) survey, constructed by the CAMIRA
(Cluster-finding Algorithm based on Multi-band Identification of Red-sequence 
gAlaxies) algorithm \citep{Oguri:2014,Oguri+:2017}.  
Our work extends the previous study of \citet{Branchini+:2017} 
by adding high-redshift clusters and studying the redshift dependence 
of cross correlations between clusters and the UGRB. 
Our work using the CAMIRA cluster catalogue is also complementary to 
that of \citet{Cuoco+:2017} in which galaxy catalogues have been used. 
For instance, the CAMIRA cluster catalogue has almost homogeneous redshift 
distribution, whereas the galaxy catalogues that have been used in 
\citet{Cuoco+:2017} contain the relatively small number of galaxies at 
$0.6 < z < 1$. Furthermore, the difference in physical properties of 
these catalogues may result in different cross-correlation signals 
possibly originating from different contributions from star-forming 
galaxies and blazers. Put another way, the comparison of cross-correlation
signals for galaxy and cluster catalogues may tell us which of galaxy or 
galaxy cluster are more likely to be associated with $\gamma$-ray emitters. 

In our analysis, we divide the CAMIRA clusters into two redshift bins,
$0.1 < z < 0.6$ and $0.6 < z < 1.1$, 
and perform a stacking analysis and  cross-correlation analysis 
with the UGRB map of 1--100~GeV and the clusters with 
three redshift ranges of $0.1 < z < 1.1$, $0.1 < z < 0.6$, and $0.6 < z < 1.1$. 
We consider a simple model for astronomical $\gamma$-ray emitters~
(blazars, star-forming galaxies and radio galaxies) 
to estimate the contributions to the UGRB, 
and compare the measured cross correlation and the model prediction. 

This paper is organized as follows. 
We first describe the CAMIRA cluster catalogue, the \emph{Fermi}-LAT data, 
and how to construct the UGRB map in our analysis in Section~\ref{sec:data}.
In Section~\ref{sec:stack}, we summarize the method of the stacking analysis 
and the brief result. 
The method of the cross-correlation analysis and 
the result are presented in Section~\ref{sec:cca}. 
In Section~\ref{sec:implication}, we introduce a simple model for 
$\gamma$-ray emitters~(blazars, star-forming galaxies and radio galaxies) 
to compare with the measured cross correlation. 
We summarize our results and conclude this paper in Section~\ref{sec:conclusion}. 


%
\section{Data}
\label{sec:data}
%

\subsection{HSC photometric sample and CAMIRA cluster}
\label{ssec:hsc_photo}
The HSC survey is an imaging survey which has started observing in March 2014, and earns a large number of galaxies with wide field camera installed on the prime focus of the Subaru telescope \citep{Miyazaki2018Camera, 2018PASJ...70S...2K}. The HSC survey equips five broad filter bands and three narrow filter bands in the visible optical wavelength (Kawanomoto et al. in prep.) and aims at measuring shapes and distances of galaxies down to the limiting depth of $i\sim 26$ \citep{Aihara2018Survey,Aihara2018DR1}. With the help of an on-site quality assurance system~\citep{2018PASJ...70S...3F}, the HSC can achieve an unprecedentedly accurate photometry and shape measurement which enables us to conduct a secure weak lensing analysis. Although the HSC survey has already made the first public data release in 2017 \citep{Aihara2018DR1}, we use an internal \textit{S16A} data which reaches down to 26.4 in $i$-band PSF (Point Spread Function) magnitude and covers more than 200 square degrees of the sky spread over six fields in the northern hemisphere. The fields used in this paper are named 
\texttt{GAMA09H} (+09h, +1d30min), 
\texttt{GAMA15H} (+15h, +0d),
\texttt{HECTOMAP} (+16h12min, +43d30min), 
\texttt{VVDS} (+22h24min, +01d), 
\texttt{WIDE12H} (+12h, +00d)
and 
\texttt{XMM} (+02h15min, -04d), where the coordinates in the parenthesis are the central sky positions for the \textit{S16A} data set in the equatorial coordinate. As the HSC survey proceeds, these areas will eventually be connected to each other and will be three distinct sky areas amounting to 1400 square degrees coverage.

Each observation pointing is divided into 4 exposures for $g$- and $r$-bands, and 6 for $i$-, $z$-, and $y$-bands with a large dithering step of $\sim$$0.6$~deg, to fill the gaps between CCDs and to obtain a continuous image with roughly uniform depth over the entire area. The expected 5$\sigma$ limiting magnitudes for the $2''$ diameter aperture are 26.5, 26.1, 25.9, 25.1, and 24.4 for $g$, $r$, $i$, $z$, and $y$-bands, respectively \citep{Aihara2018Survey}. The observed images are reduced by the processing pipeline called \texttt{hscpipe} \citep{Bosch+:2017}, which is developed as a part of the LSST (Large Synoptic Survey Telescope) pipeline \citep{Ivezic+:2008, Axelrod+:2010, Juric+:2015}. The photometry and astrometry are calibrated in comparison with the Pan-STARRS1 3$\pi$ catalogue \citep{Schlafly+:2012, Tonry+:2012, Magnier+:2013}, which is totally overlapped with the HSC survey footprint with similar filter response functions to the HSC. 
The flux measurement of the current \texttt{hscpipe} 
particularly in the vicinity of cluster centers is not accurate.  
One of the most serious reason is that the objects are too close to
each other and significant overlaps makes it difficult to resolve
individual component on the images, \textit{deblending}
\citep{Bosch+:2017}. To get around this issue, \texttt{hscpipe} 
also provides the PSF-matched aperture phtometry on parent images
before deblending, which was used in cluster finding described below.

CAMIRA is a cluster finding algorithm based on the red-sequence
galaxies first applied to the Sloan Digital Sky Survey (SDSS) galaxies
\citep{Oguri:2014}. 
\citet{Oguri+:2017} applied the slightly upgraded algorithm to the 
HSC S16A data to construct a catalogue of $\sim 1900$ clusters from
$\sim 230$~deg$^2$ of the sky over the redshift
$0.1<z<1.1$ with almost uniform completeness and purity. 
The cluster finding procedure is as follow.
\begin{enumerate}
\item{}As the first step, CAMIRA applies specific colour cuts to a
spectroscopic redshift-matched catalogue
in the redshift-colour diagram to remove the obvious blue  galaxies,
which makes latter procedure more efficient and secure.
Although the colour cuts by eyeballing might induce some
artificial effects, those galaxies are used only for calibrating the
colour of the red-sequence galaxies and this colour-cut is not applied
in cluster finding itself. The spectroscopic galaxies are used to
calibrate and improve the accuracy of the stellar population 
synthesis (SPS) model as described below.
\item{}Then, CAMIRA derives the likelihood of a galaxy being on the
red-sequence as a function of redshift, by fitting the colours of 
the HSC galaxy to those predicted by a stellar
SPS model of \citet{BC03}. The SPS model is calibrated and improved 
by comparing model predictions with observed galaxy colours of 
spectroscopic galaxies in the HSC footprint.
Not only the photometric redshift but also the richness
parameter is also computed with which a cluster candidate is
identified as a peak of the richness map.
\item{} For each identified cluster, the Brightest Cluster Galaxy (BCG) is assigned as
  a bright galaxy near the richness peak. The photometric redshift and
  richness are iteratively updated until the result converges. As a
  result, the accuracy of the photometric redshift reaches 1~per~cent out to $z\sim 1$.
\end{enumerate}

In this paper, we use 4948 clusters from $\sim
230$~deg$^2$ HSC six wide fields with richness 
$\hat{N}_{\rm mem}>10$, which roughly corresponds to mass of 
$M_{\rm 200m}>10^{13.5} M_{\odot}h^{-1}$ with a long tail to lower
masses \citep{Murata+:2018}.
Although only clusters with $\hat{N}_{\rm mem}>15$ are published in \citet{Oguri+:2017},
we push down the richness limit in order to have a larger number of clusters to be 
stacked. For less massive clusters, $10<\hat{N}_{\rm mem} < 15$, the centre of cluster 
or the membership of galaxies include larger uncertainties compared to those for massive clusters.
However, the signal we are searching for is projected along the line of sight, and the
coarse angular resolution of \emph{Fermi} data is much larger than the centric uncertainties of the CAMIRA cluster catalogue.
Therefore we use these less massive clusters, together with the published CAMIRA 
clusters with $\hat{N}_{\rm mem}>15$, 
to increase the detection significance as much as possible.
We remove the 232 clusters within the point source mask that
is described in Section~\ref{subsection:FermiDataSelection}.
In addition, 255 clusters near the edge of the fields are removed
because, when we stack the $\gamma$-ray images, they bring a deficit on
the image. 
We divide the cluster sample into two different redshift ranges according to 
the photometric redshift of clusters: 1942 clusters at 
$0.1<z_{\rm cl}<0.6$ and 2519 clusters at $0.6<z_{\rm cl}<1.1$.
Sky distribution of clusters in each redshift range is shown in Figure~\ref{fig:camiramap}, and redshift distribution is shown in Figure~\ref{fig:reddis}.

\iffigure
\begin{figure*}
  \begin{center}
    \begin{tabular}{cc}
      \includegraphics[width=0.5\linewidth]{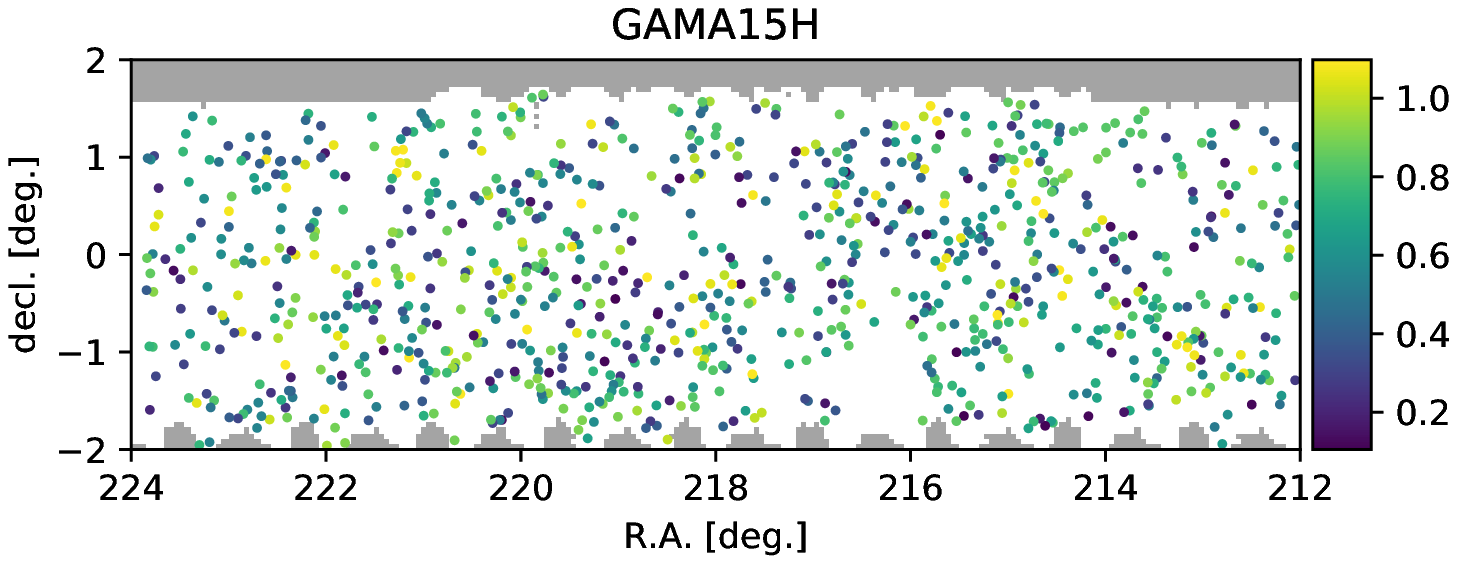}&
      \includegraphics[width=0.5\linewidth]{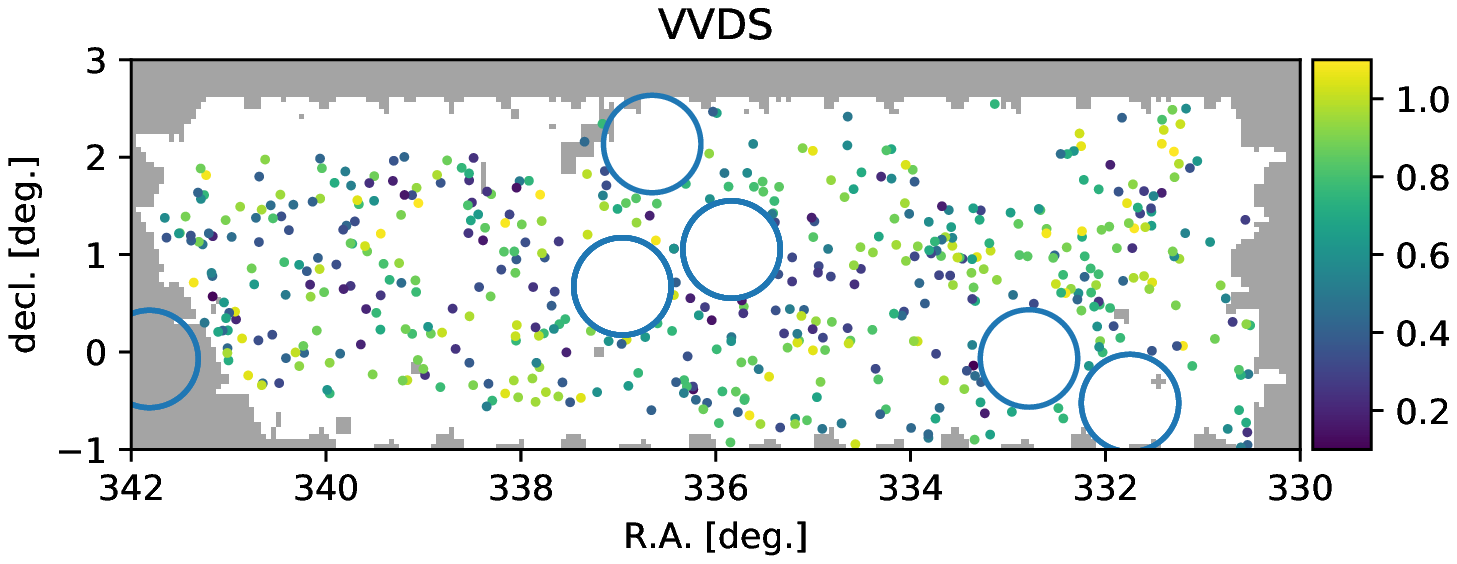}\\
      \includegraphics[width=0.5\linewidth]{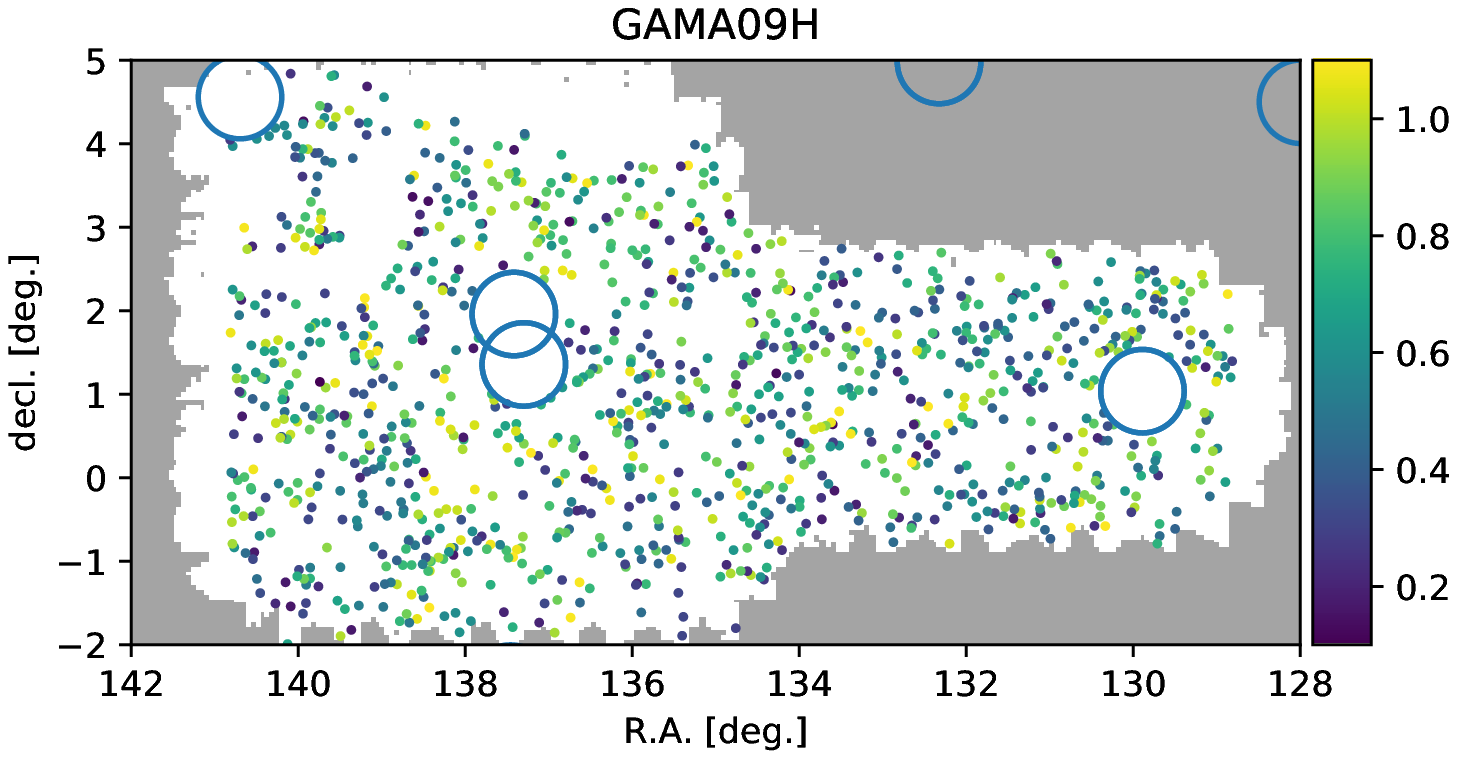}&
      \includegraphics[width=0.5\linewidth]{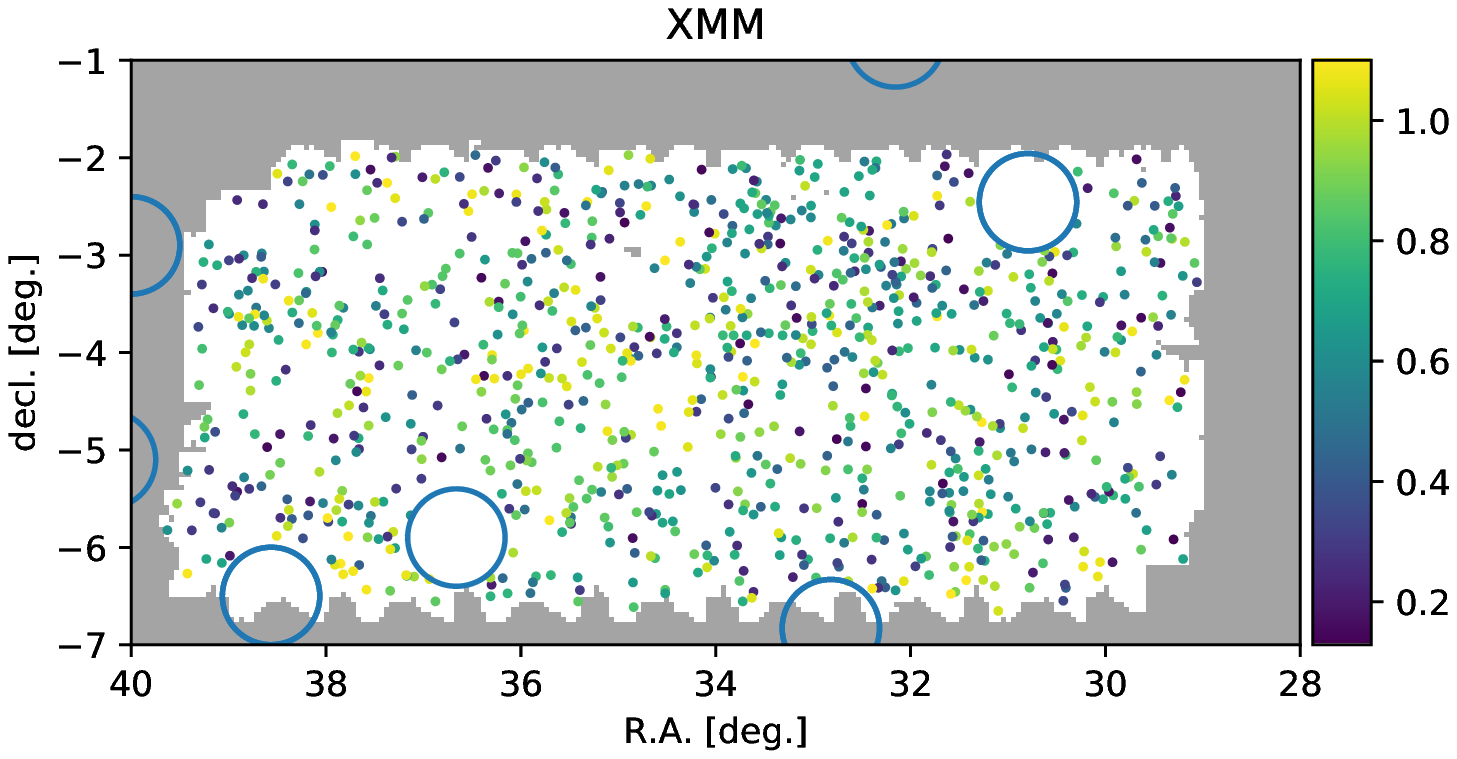}\\
      \includegraphics[width=0.5\linewidth]{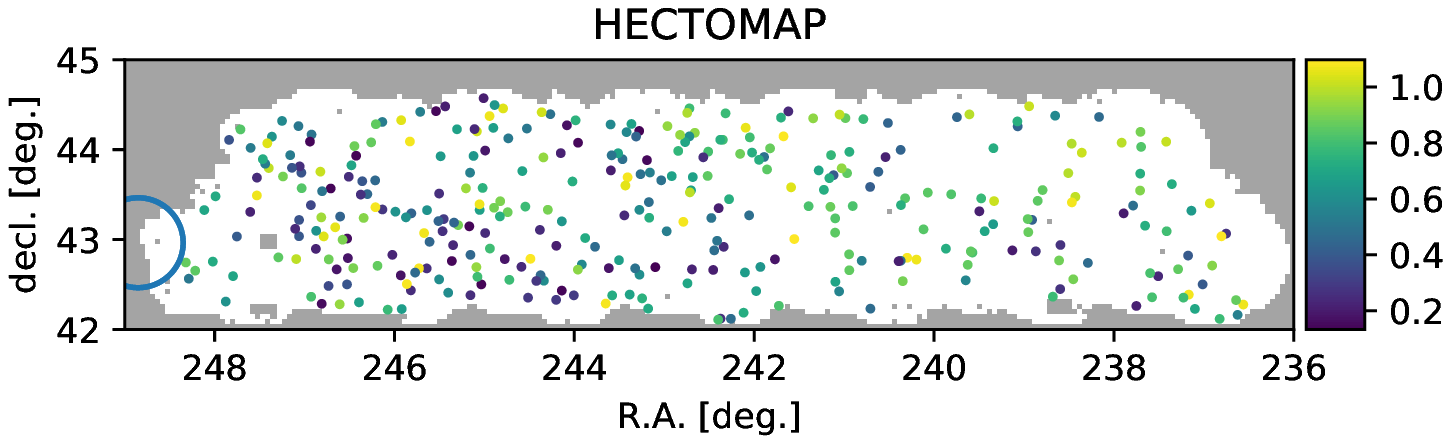}&
      \includegraphics[width=0.5\linewidth]{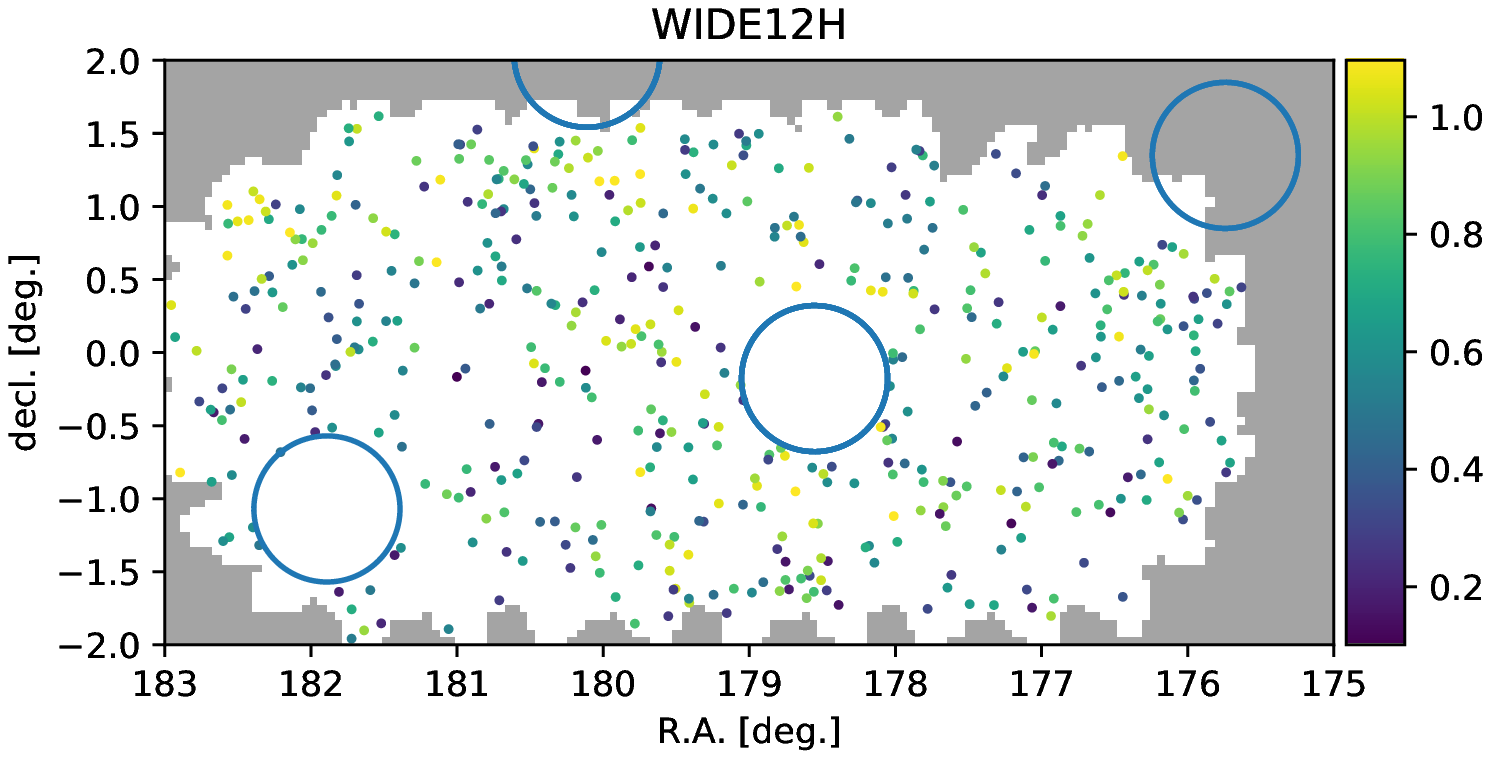}\\
    \end{tabular}
  \end{center}
 \caption{
   The sky distribution of the CAMIRA clusters for 6 patches used in this work.
    Points show centre positions of the clusters and colour level represents cluster's redshift.
    Circles are masked regions with radii of $0.5^{\circ}$ from point-source centre~(also see Section \ref{ssec:ugrb}). 
    In shaded regions, the clusters are not identified either because
    there is no observation or the data is shallow.
    }
 \label{fig:camiramap}
\end{figure*}
\fi

\iffigure
\begin{figure}
 \begin{center}
  \includegraphics[width=1\linewidth]{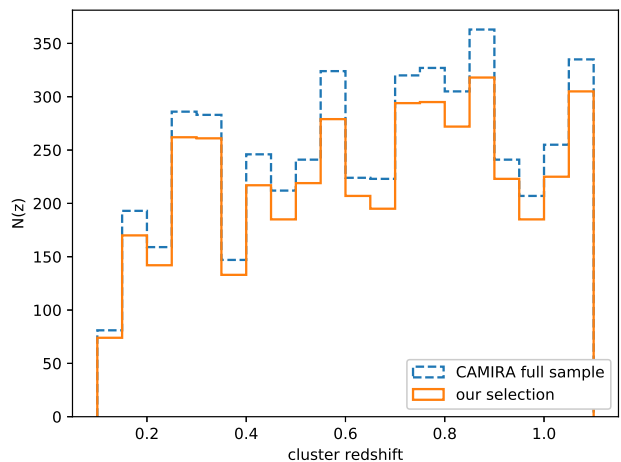}
 \end{center}
 \caption{
      The redshift distribution of CAMIRA clusters in HSC S16A. The comparison 
      with the full sample suggests that our sample selection due to the edge and 
      point mask does not significantly bias the distribution.
      }
 \label{fig:reddis}
\end{figure}
\fi

\subsection{\emph{Fermi}-LAT data reduction}
\label{subsection:FermiDataSelection}
The $\gamma$-ray analysis pipeline proceeds similarly to what is described in~\cite{2018arXiv180210257S}, and we refer the reader to that work for further details. In summary, we analyse $\sim 7$ years (from August 4, 2008 to September 4, 2015) of Pass~8 {\tt ULTRACLEANVETO} photons and select $\gamma$-ray events  in the energy range $E=[1,100]$ GeV. We impose standard selection cuts on the $\gamma$-ray data, removing events entering at zenith angles larger than 90$^\circ$ to reduce cosmic-ray contamination. Moreover, the photon data was filtered by removing time periods when the instrument was not in sky-survey mode. To produce flux maps in our regions of interest~(ROI) we use the \emph{Fermi} Science Tools v10r0p5 software package and the instrument response functions (IRFs), {\tt P8R2\_ULTRACLEANVETO\_V6}.

As described in Appendix A of~\cite{2018arXiv180210257S}, we analyse $\gamma$-rays within square regions of size $20^{\circ} \times 20^{\circ}$ around each of the HSC fields of view. To reduce the computation time of our pipeline in the ROIs we further divided each $20^\circ \times 20^\circ$ region into four $12^\circ \times 12^\circ$ contiguous patches with overlapping boundaries. The measured $\gamma$-rays at each patch were modeled by a linear combination of diffuse $\gamma$-ray background and foreground models as well as lists of  $\gamma$-ray point sources present in the ROIs.

In particular, we employ two different kinds of interstellar emission models (IEMs) in order to estimate the systematic uncertainties introduced by the choice of IEM. The first IEM, included in our baseline model, corresponds to the standard LAT diffuse emission model, {\tt gll\_iem\_v06.fits}, this is the model routinely used in Pass~8 analyses. As alternative IEMs,  we consider three different diffuse models produced with the GALPROP\footnote{See \url{http://galprop.stanford.edu}.} cosmic-ray propagation code. These alternative IEMs (called Models A, B and C) were introduced in~\cite{Ackermann+:2015} and encompass a very wide range of the systematics associated with this kind of analysis. As such, they provide a test in rigor comparable to that performed by the \emph{Fermi} team. 

The fit started with a sky model that includes all point-like and extended LAT sources listed in the 3FGL~\citep{3FGL} catalogue as well as a list of new point sources found in~\cite{2018arXiv180210257S} (see Table III in the appendix section). In our baseline model the Galactic diffuse emission was modeled by the standard LAT diffuse emission model, and as a proxy for the residual background and extragalactic $\gamma$-ray radiation we used the isotropic template given by model, \texttt{iso\_P8R2\_ULTRACLEANVETO\_V6\_v06.txt}\footnote{http://fermi.gsfc.nasa.gov/ssc/}. Best-fit spectral parameters were obtained for all free sources within our $12^\circ \times 12^\circ$ ROIs. To obtain convergence, all the fits were performed hierarchically; freeing first the normalization of the sources with the highest intensities followed by the lower ones within the ROIs. The fitting consecutively restarts from the updated best-fit models and repeats the same procedure this time for the spectral shape parameters.

The EGB in the ROIs were obtained by subtracting the best-fit Galactic diffuse emission model from the photon counts maps. We note that the EGB images obtained in this way could still contain some isotropic detector backgrounds. However, our analysis is able to reproduce well the EGB $\gamma$-rays derived by the \emph{Fermi}-LAT team (see, e.g. Ref.~\cite{Shirasaki:2016kol} for an example of this method). This shows 
that most of the detector cosmic-ray induced backgrounds are safely removed by our conservative photon selection filters.

\subsection{UGRB map}
\label{ssec:ugrb}

\iffigure
\begin{figure*}
  \begin{center}
   \begin{tabular}{cc}
   \includegraphics[width=32mm, angle=-90]{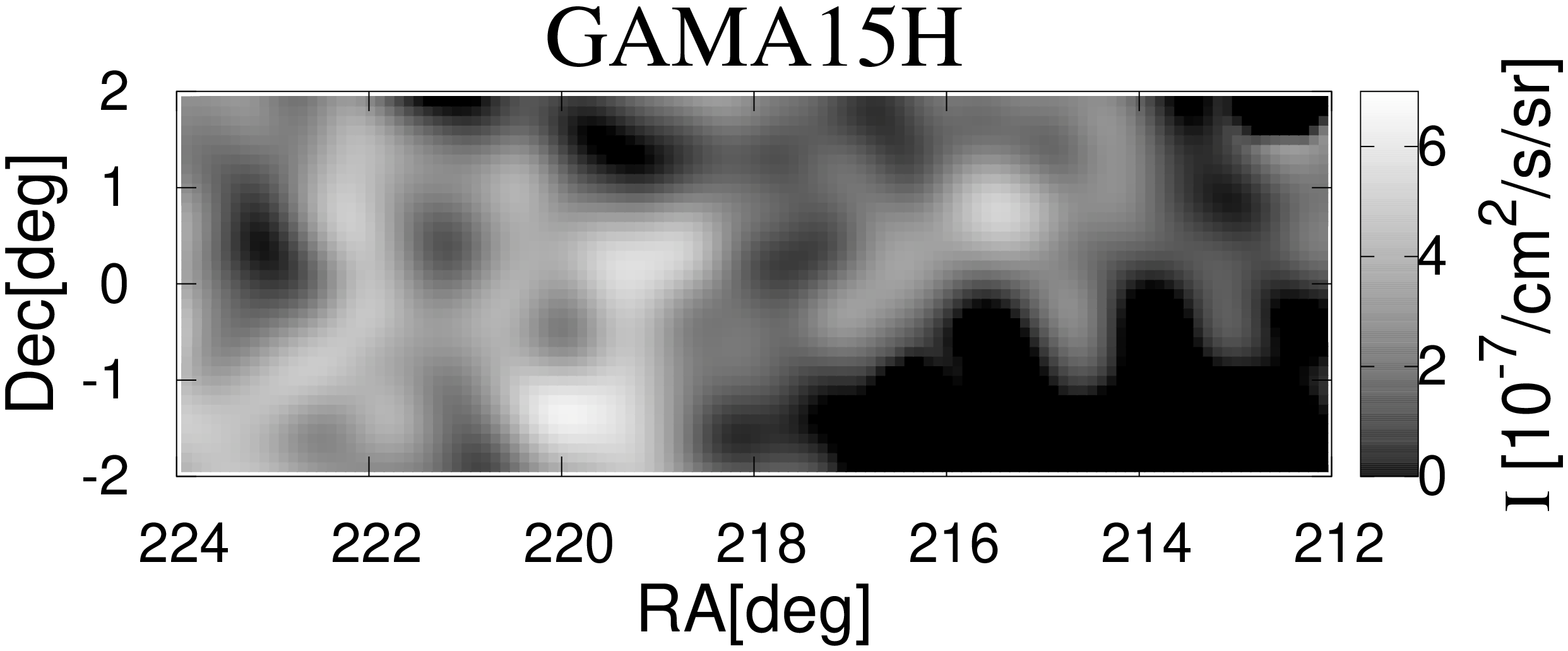}&
   \includegraphics[width=30mm, angle=-90]{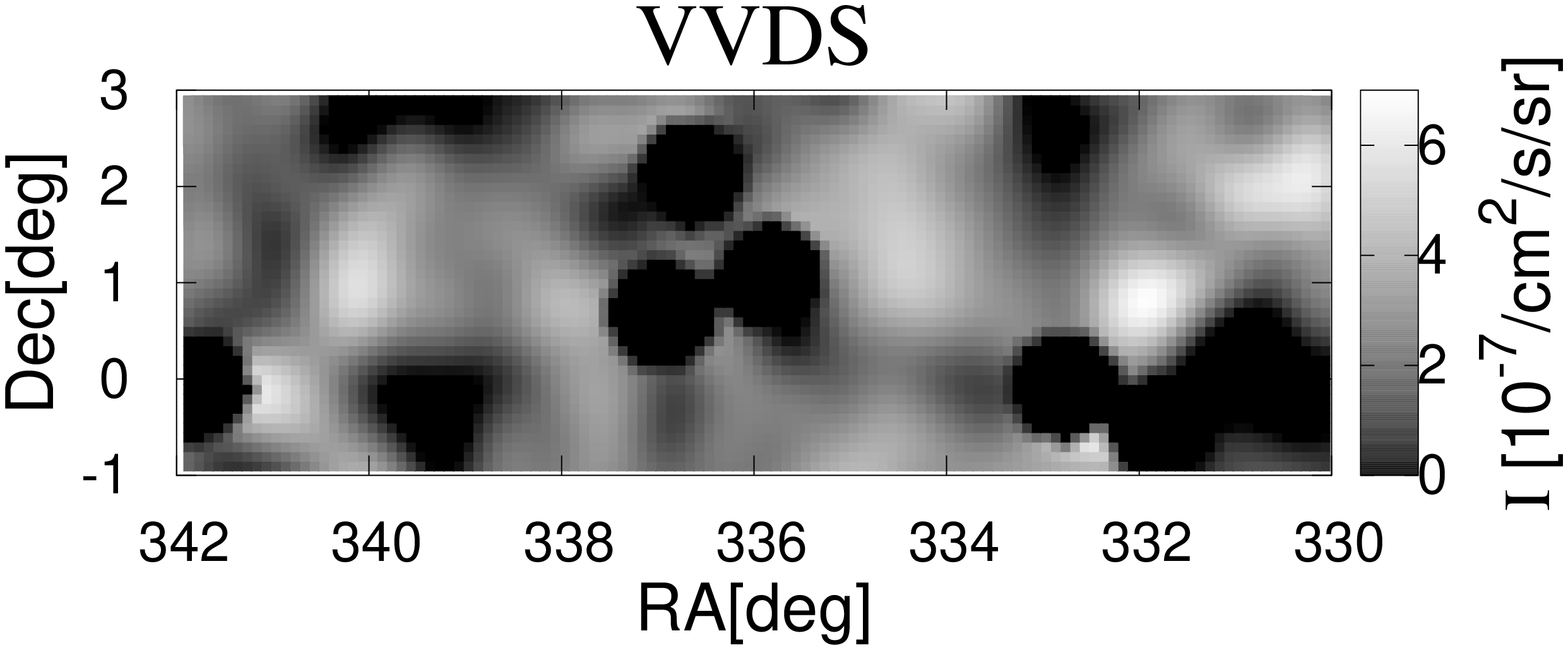}\\
   \includegraphics[width=40mm, angle=-90]{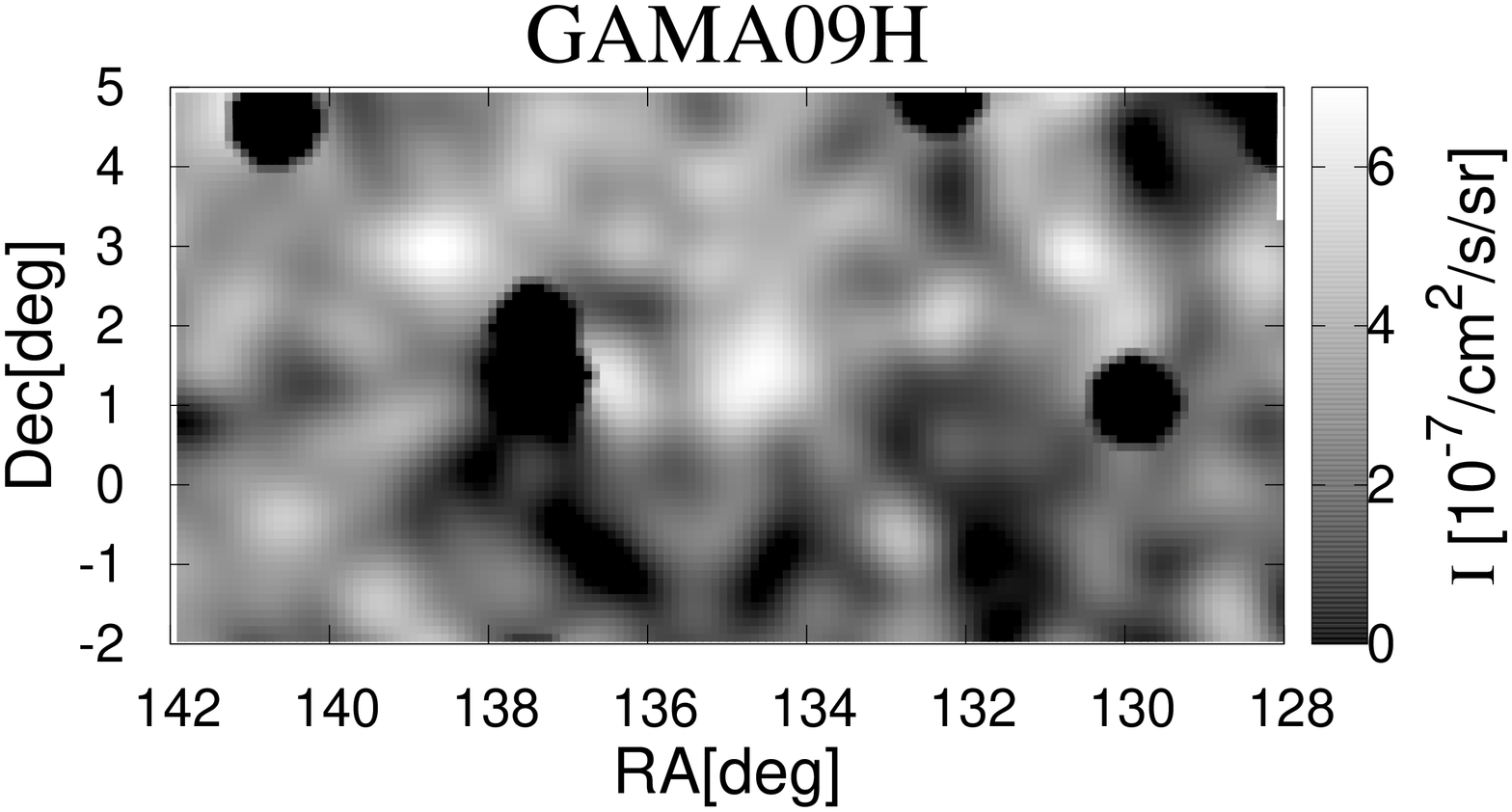}&
   \includegraphics[width=38mm, angle=-90]{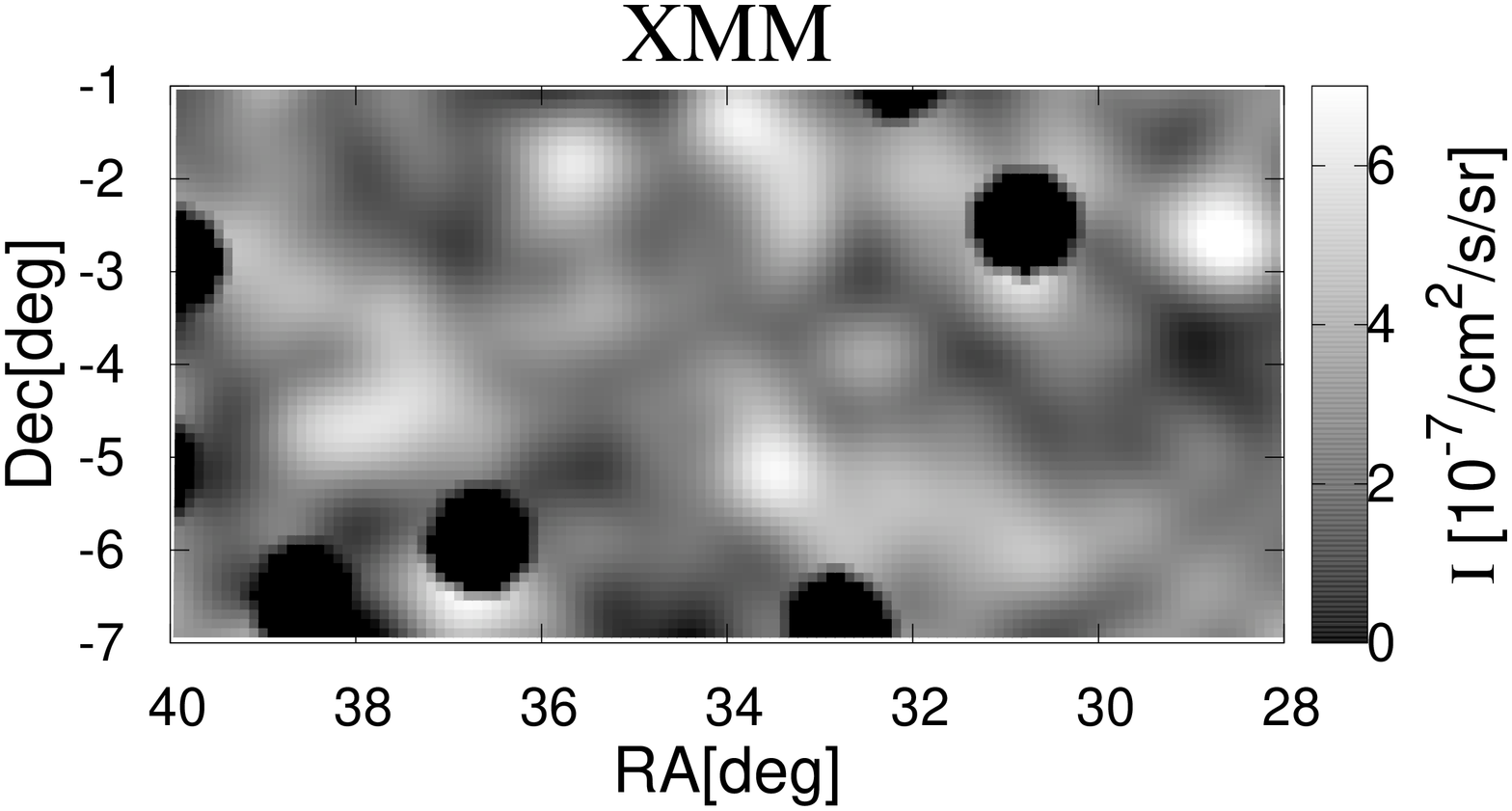}\\
   \includegraphics[width=25mm, angle=-90]{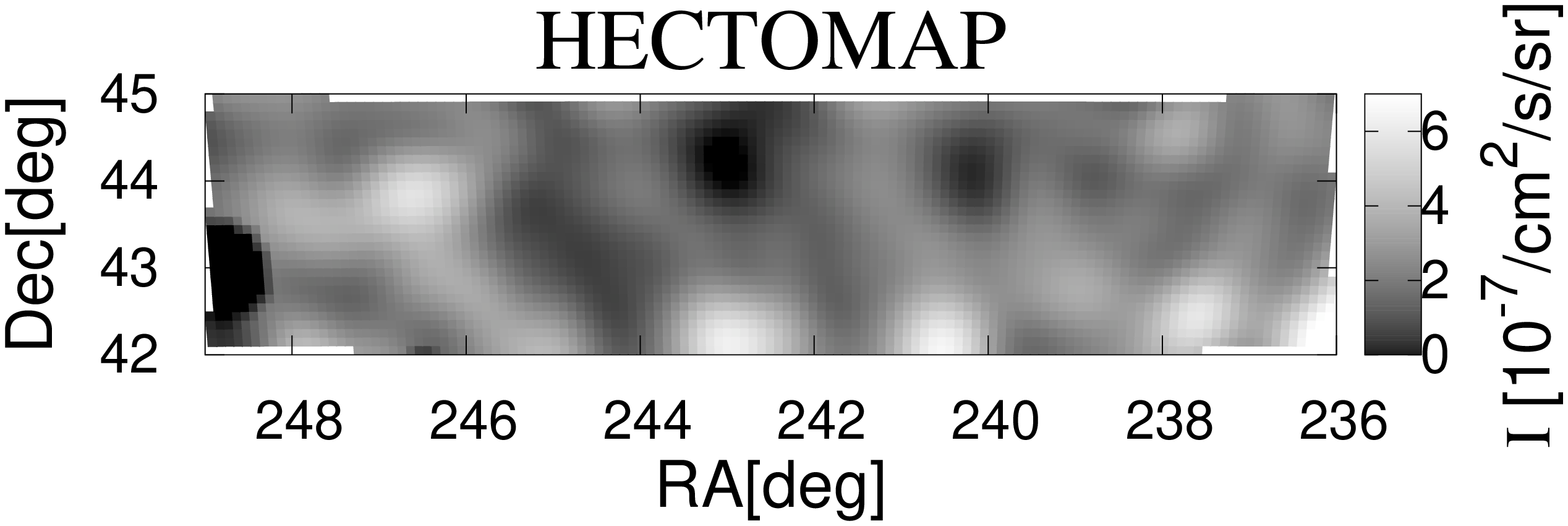}&
   \includegraphics[width=38mm, angle=-90]{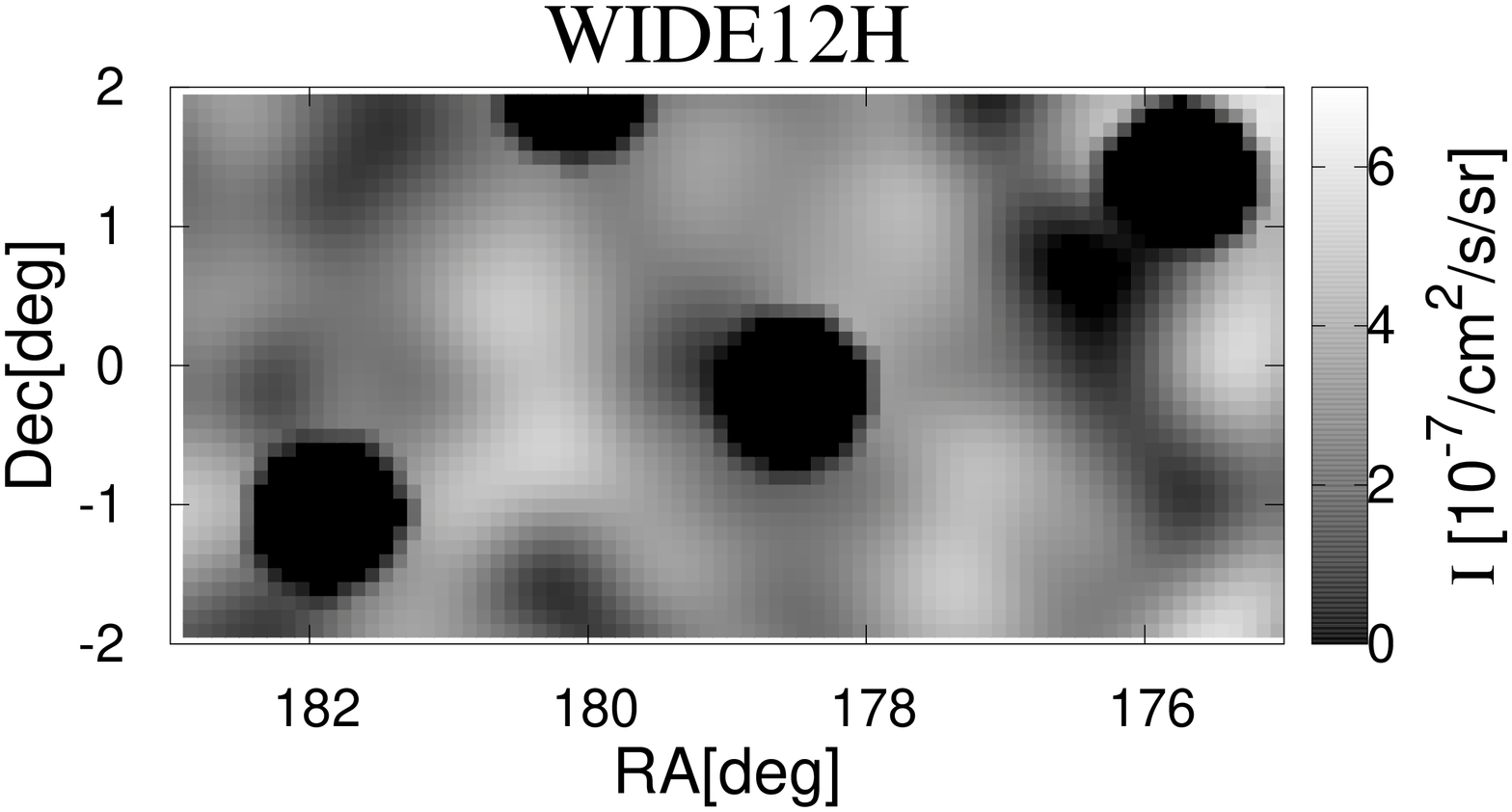}\\
  \end{tabular}
 \end{center}
 \caption{
    UGRB maps corresponding to CAMIRA-cluster regions~(Figure~\ref{fig:camiramap}) in our analysis.  The colour bar is an amplitude of the number intensity for $\gamma$-ray photons.  In this figure, the UGRB maps are constructed from \emph{Fermi} $\gamma$-ray data of 1-100 GeV by masking point-source regions~(circular black shaded areas),   subtracting the Galactic $\gamma$-ray emission for the baseline model and applying a Gaussian smoothing of $0.5^{\circ}$. Note that only in this figure, we do not only apply a smoothing scale of $0.5^{\circ}$ below 10 GeV but also higher than 10 GeV. 
    }   
 \label{fig:fermimap}
\end{figure*}
\fi

In our analysis, we use the \emph{Fermi} $\gamma$-ray data of 1-100 GeV.  To subtract the diffuse Galactic $\gamma$-ray emission, four foreground models are applied: baseline model, Model A, B, and C \citep{Ackermann+:2015}.  The mean number intensities of the $\gamma$-ray photons for the UGRB applied each foreground model are 2.57, 2.04, 2.24 and 2.45$~(10^{-7}{\rm cm^{-2}s^{-1}sr^{-1}})$  for baseline model, Model A, B and C, respectively.  We also subtract emission from resolved point sources listed in the 3FGL catalogue~\citep{3FGL} and new point-source candidates~\citep{shirasaki+:2018}. 
To be more conservative, we mask circular regions of radius $0.5^{\circ}$ around the point sources to avoid a possible overcorrection for the $\gamma$-ray fluxes.

Since the mean number intensity of $\gamma$-ray per pixel is small, $\sim 0.1$, we apply a Gaussian smoothing filter on the UGRB map to relax the shot noise effect. The smoothing scales depend on the energy of $\gamma$-ray observed and we take 0.5$^{\circ}$ for below $10{\rm GeV}$, and 0.2$^{\circ}$ otherwise, which roughly correspond to the size of PSF, which again depends on the energy scale of $\gamma$-ray \citep{Ackermann+:2012}.
Figure~\ref{fig:fermimap} shows the processed $\gamma$-ray map in six different fields where all the $\gamma$-ray photon is integrated over 1-100 GeV.

\section{STACKING ANALYSIS} 
\label{sec:stack} 

In this section, we perform the stacking analysis between the UGRB map and CAMIRA clusters. 
The stacking analysis is useful to visually find a correlation signal and is 
basically equivalent to the cross-correlation analysis which we describe 
in section~\ref{sec:cca}. 
This analysis can increase a signal-to-noise~(SN) ratio
by combining multiple images.
In this section, we describe the method of stacking analysis and show the result. 

First we select a $4^{\circ} \times 4^{\circ}$ square image of the UGRB, centred at each cluster position, which is defined as the location of the most probable BCG among the member galaxies\citep{Oguri+:2017}.  After randomly rotating the images, we stack all images to obtain the average image of $\gamma$-ray around the cluster, we call it as stacked image at ``cluster'', 
\begin{equation}
\label{eq:stack}
   I_{\rm stack}^{\rm clu}(\bs{\theta})
   =
   \frac{1}{N_{\rm clu}\overline{I_{\gamma}}}
   \sum_i^{\rm N_{clu}}
   {\mathcal G}(\bs{\phi}_i)I_{\gamma} (\bs{\Theta}_{i}-\bs{\Theta}_{i,\rm clu}),
\end{equation}
where $\bs{\theta}$ is a separation angle from cluster centre,  $N_{\rm clu}$ is the number of clusters to be stacked,  $I_{\gamma, i}(\bs{\Theta}_{i} - \bs{\Theta}_{i,\rm clu})$ is the number intensity of $\gamma$-ray for around the $i$-th cluster centred at  $ \bs{\Theta}_{i,\rm clu}$ and  
$\overline{I_{\gamma}}$ is the averaged $I_{\gamma}$ over all ROIs. The rotation operator ${\mathcal G}$ is defined in the local coordinate and its argument $\bs{\phi}_i$ is randomly selected from $[0,2\pi]$. The random rotation of each map can reduce the effect of  large scale anisotropies due to the imperfect subtraction of Galactic foreground component. 

The stacked image at ``cluster" might include not only the $\gamma$-ray light from clusters identified in the CAMIRA catalogue but also the background or foreground $\gamma$-ray not associated to the clusters. To subtract any components not related to clusters from the stacked image, we also make random stacked images through the similar procedure, in which we select random positions as the centre of the images for stacking, instead of the CAMIRA cluster position: i.e. by replacing the superscript of ``clu" with ``ran" in equation~(\ref{eq:stack}). Then, we obtain the final stacked $\gamma$-ray image  by subtracting the ``random" image from the ``cluster" one. 
\begin{equation}
\delta_{\rm stack} (\bs{\theta})
=
I_{\rm stack}^{\rm clu}(\bs{\theta}) -
I_{\rm stack}^{\rm ran}(\bs{\theta}) 
\end{equation}

\iffigure
\begin{figure}
 \begin{center}
  \begin{tabular}{cc}
   \includegraphics[width=35mm, angle=-90]{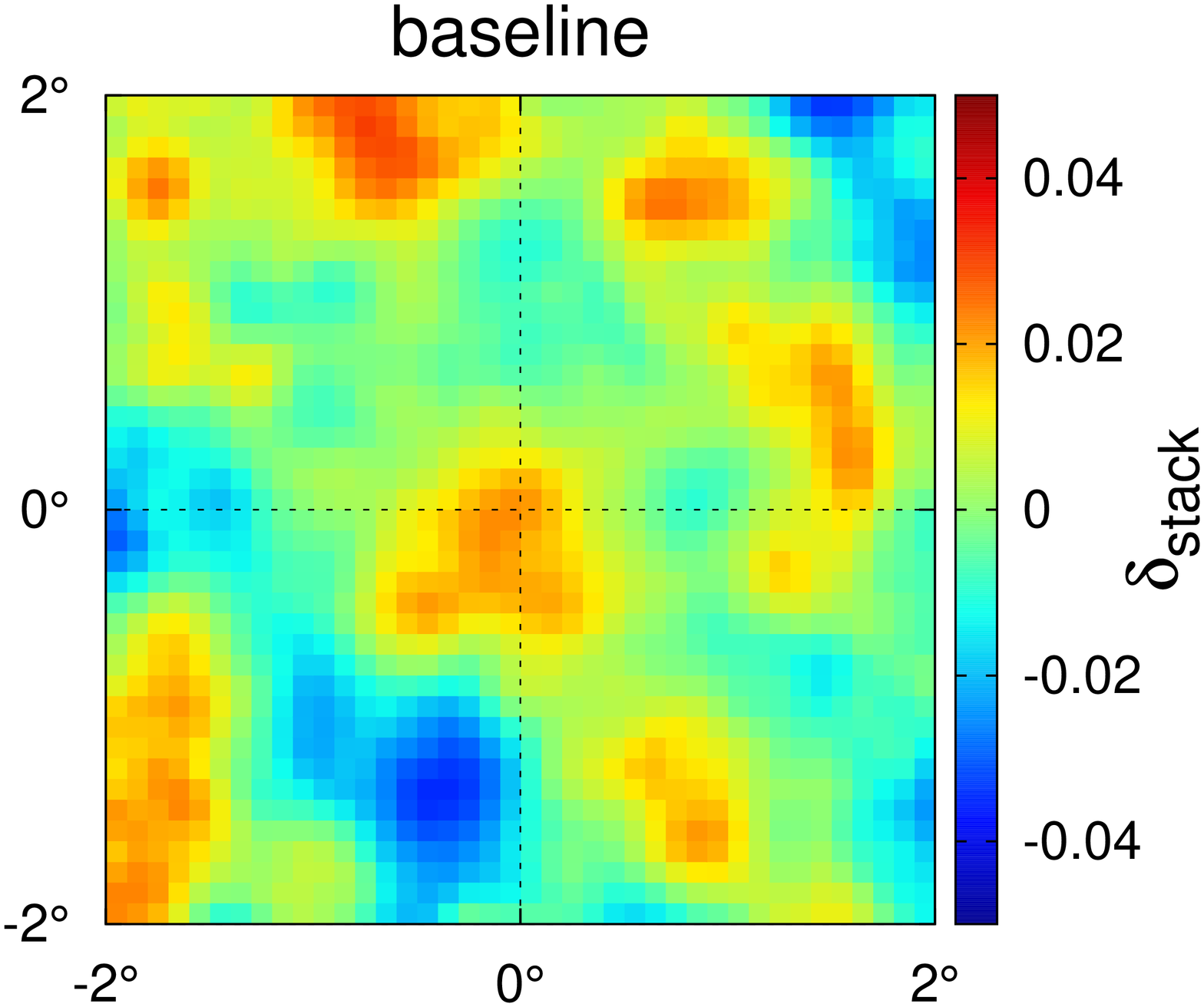}&
   \includegraphics[width=35mm, angle=-90]{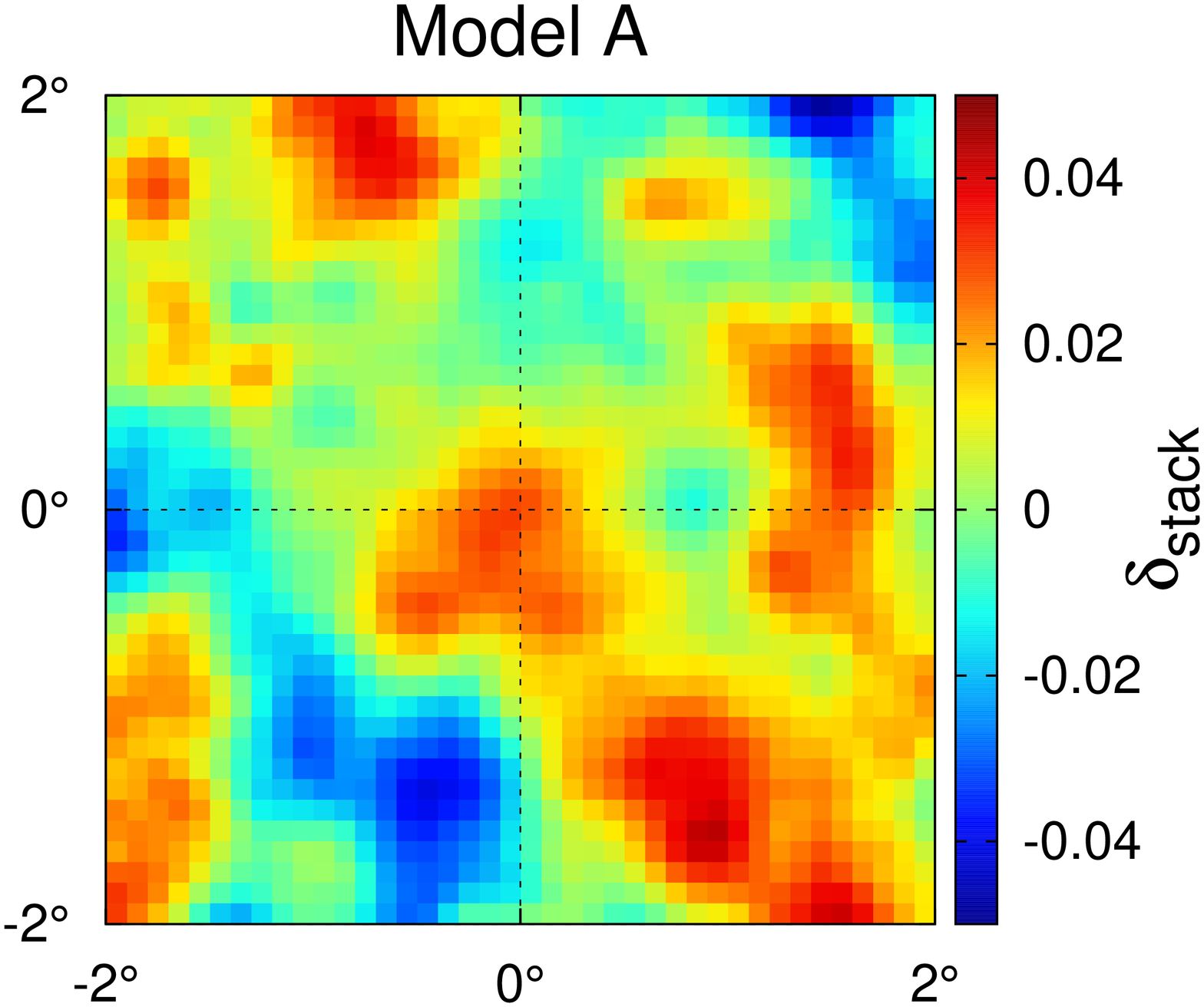}\\
   \includegraphics[width=35mm, angle=-90]{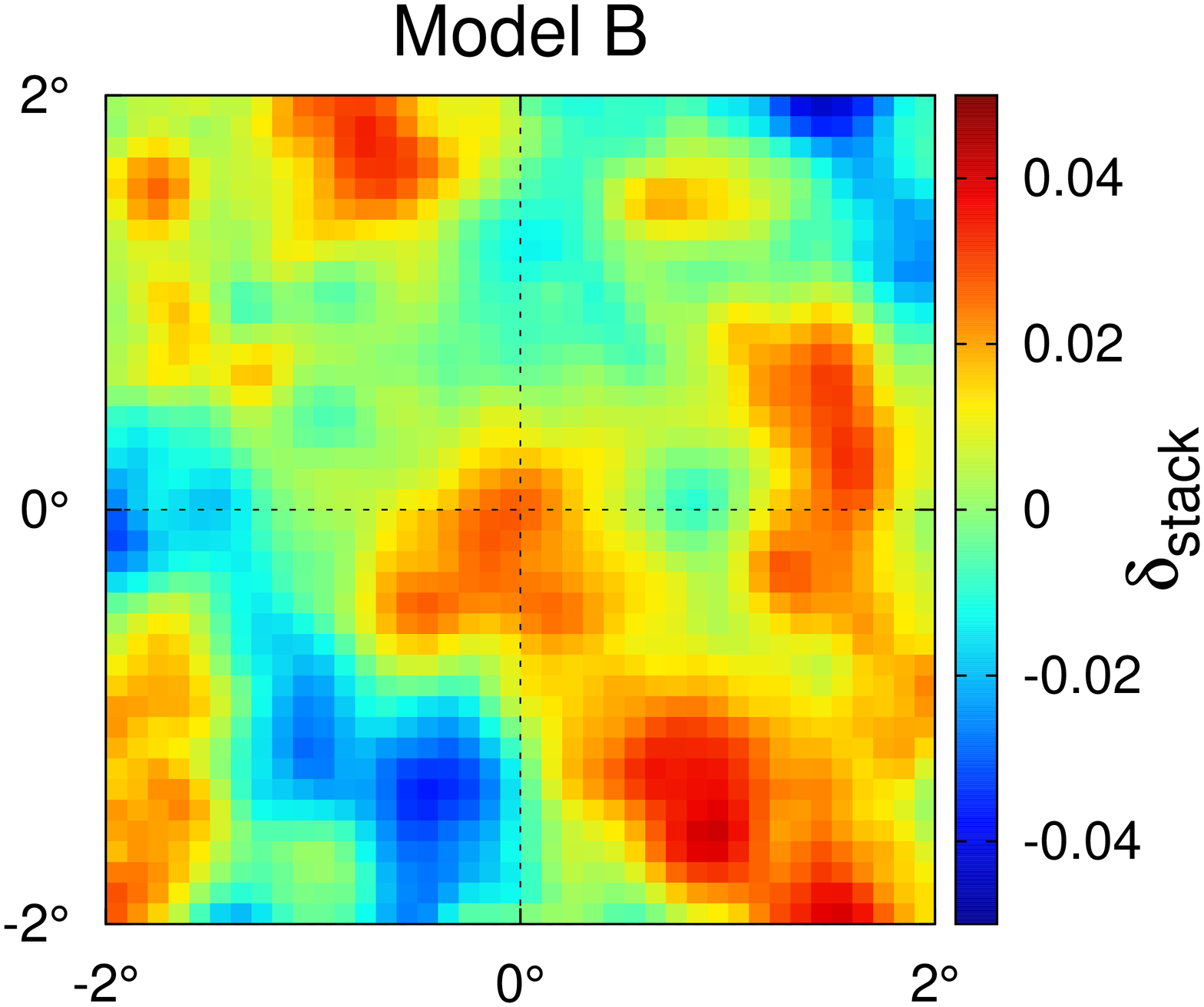}&
   \includegraphics[width=35mm, angle=-90]{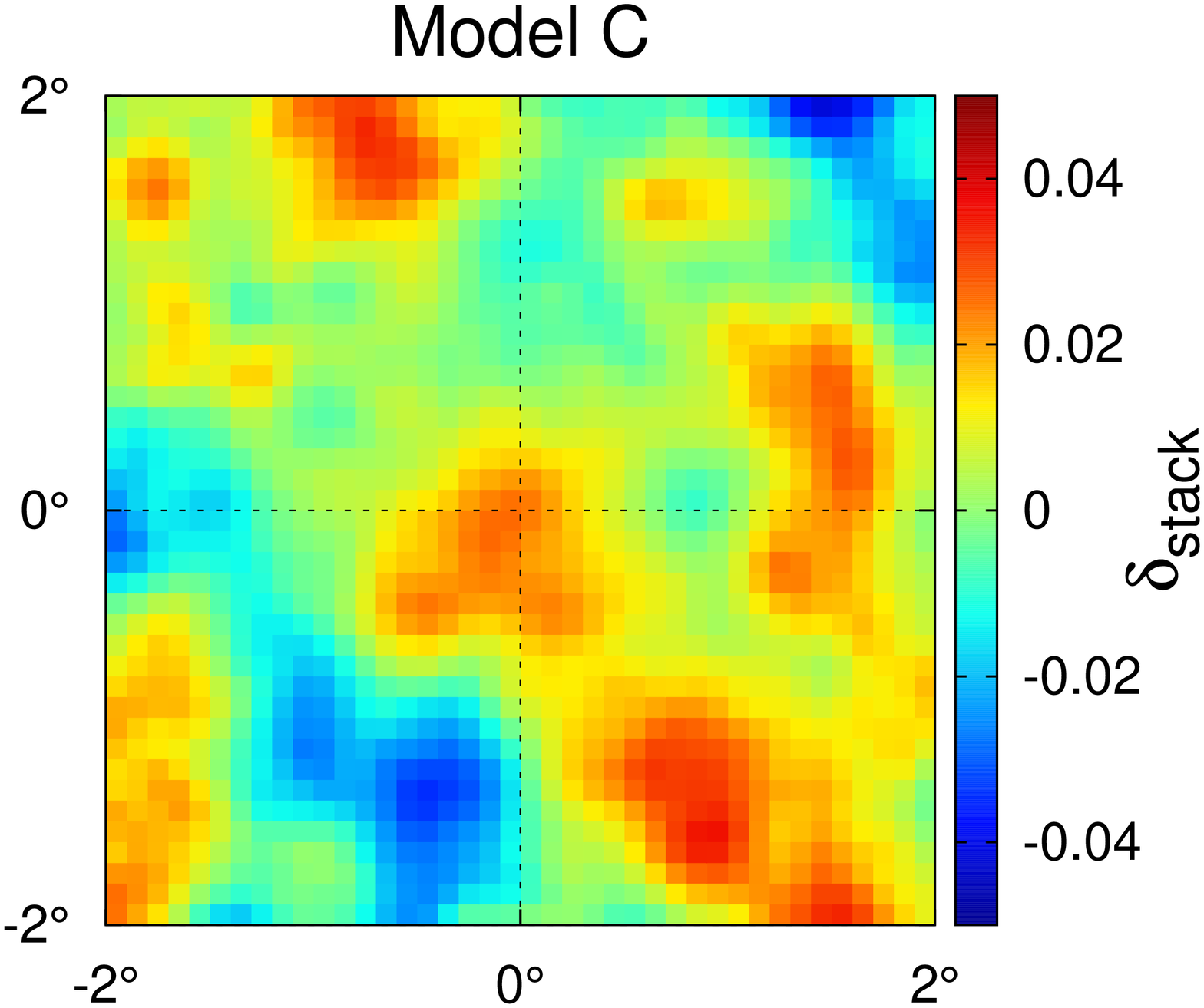}\\
  \end{tabular}
 \end{center}
 \vspace{-2mm}
 \caption{
    Stacked maps using the UGRB map for different Galactic foreground models. 
    The colour bar is an amplitude of stacked fluctuation for the $\gamma$-ray number intensity. 
    Cluster centre is located at $(0^{\circ}, 0^{\circ})$. 
    }
 \label{fig:stmap}
\end{figure}
\fi

Figure~\ref{fig:stmap} shows the final stacked image for different  Galactic foreground models.
One can clearly see the $\gamma$-ray excess at the centre region. 
This excess corresponds to a few per~cent of the 
average intensity and 
extends to $\sim 1^{\circ}$ which corresponds to about a few 10~Mpc, for typical CAMIRA cluster's redshift, $z\sim 0.6$.
This is clearly larger than the typical size of cluster.
We repeat the same analysis but limited $\gamma$-ray samples at lower energy bins, that is $<5$ GeV, the result is not changed dramatically. This means that the signal at the central region is dominated by the lower energy $\gamma$-ray photons. As we see in Section \ref{ssec:ugrb}, the PSF size depends on the $\gamma$-ray energy and thus this widely spread diffusion signature can be due to the PSF of the \emph{Fermi}-LAT, however; we do not exclude any possibilities that this diffusive signature originates from the cluster vicinity regions, like filaments or wall.
We also see in figure~\ref{fig:stmap} that the contrast of the intensity slightly depends on the foreground model, which infers the Galactic foreground models includes uncertainties at the level of the difference of the signal at central region among different models.

We repeat the above analysis for subsamples of clusters divided in two redshift ranges,  $0.1< z < 0.6$  and $0.6< z < 1.1$, which is shown in Figure~\ref{fig:st_z}. For the low-redshift cluster, a strong correlation is found around the clusters position. Moreover, the signal is spatially distributed more widely around the centre than in the case of all clusters shown in Figure~\ref{fig:stmap}. On the other hand, for the higher-redshift cluster, we only see a slight excesses at the centre. 
We note however that signal to noise ratios for the stacked map are $\sim 1$. Therefore, the stacked map has a comparable level of noise to 
the signal and the strong pattern of the map is likely due to the noise.
And also note that each $4^{\circ} \times 4^{\circ}$ image used for the stacking analysis is highly correlated with one another. The average separation of the CAMIRA clusters is $\sim 0.22~$degree and is much smaller than the image size of $\sim 4$ degrees). Therefore, some photons appear multiple times at different positions in the ``stacked'' image. For this reason, we do not conduct a quantitative discussion using the ``stacked'' image. Instead, we perform the the cross-correlation analysis in the next section.

\iffigure
\begin{figure*}
 \begin{center}
  \begin{tabular}{cc}
   \includegraphics[width=0.4\linewidth, angle=-90]{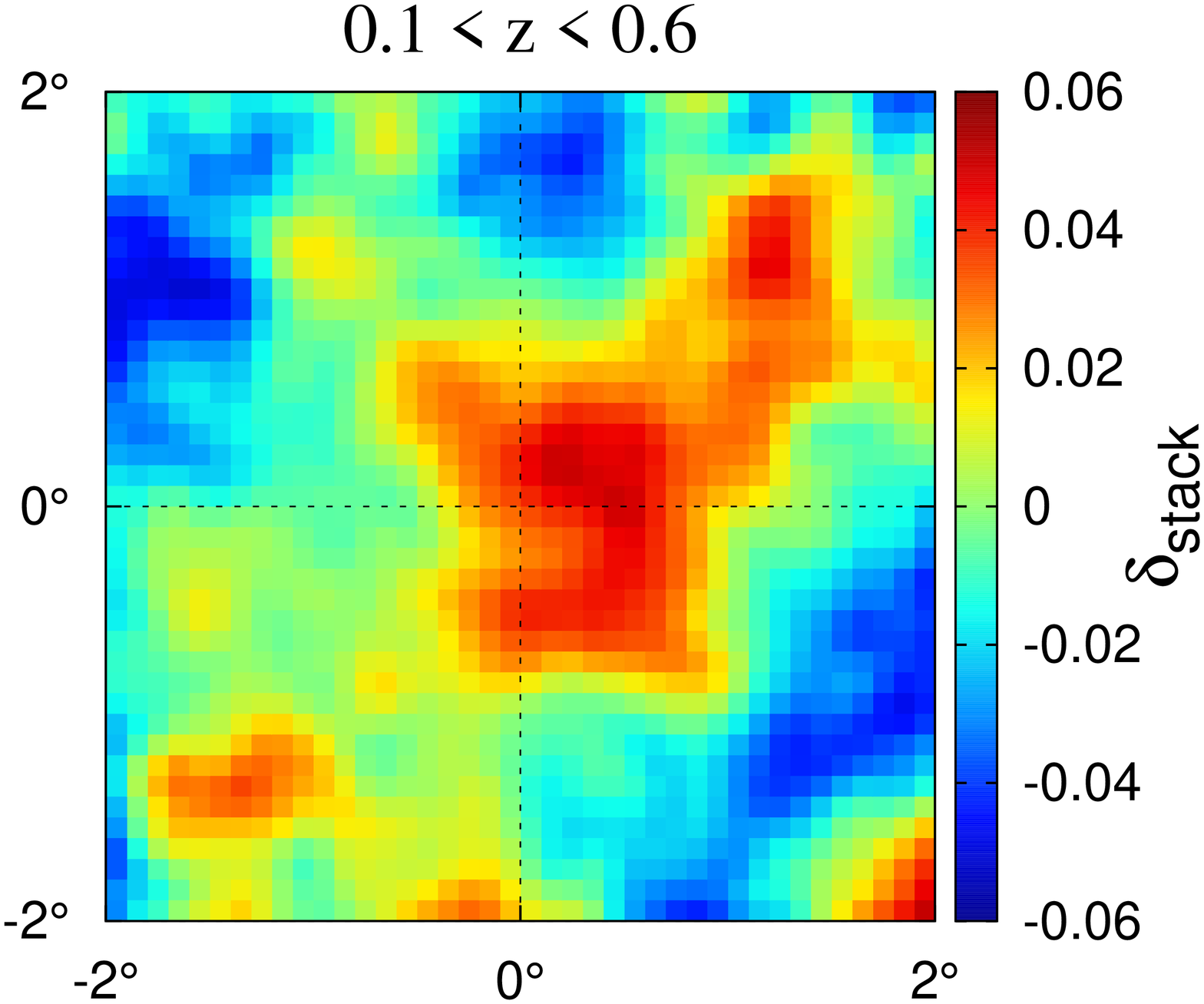}&
   \includegraphics[width=0.4\linewidth, angle=-90]{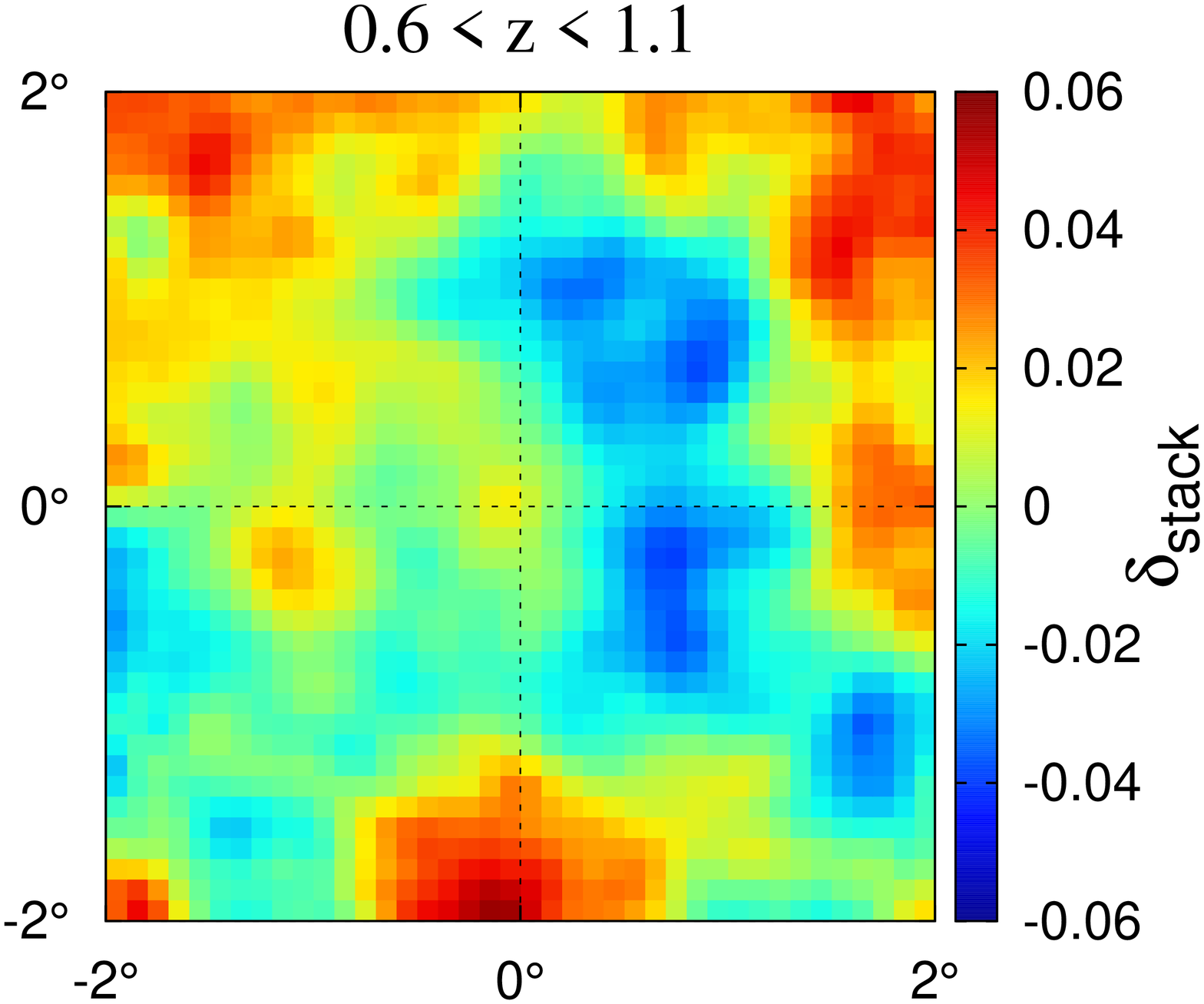}\\
  \end{tabular}
 \end{center}
 \caption{
 Same as upper left panel of Figure~\ref{fig:stmap}, but for clusters at $0.1<z<0.6$~(Left) and $0.6<z<1.1$~(Right). 
      }
 \label{fig:st_z}
\end{figure*}
\fi

\section{CROSS-CORRELATION ANALYSIS}
\label{sec:cca}
%
In this section, we perform the cross-correlation analysis between CAMIRA clusters and the UGRB map. We describe our method for the evaluation of the cross-correlation between them and then show the results. 

\subsection{correlation function}
\label{ssec:xi}
%
To evaluate the two-point cross-correlation function,
we use the Landy-Szalay estimator~\citep{Landy-Szalay:1993}.
\citet{Kerscher:2000} have shown that the Landy-Szalay estimator 
provides the most accurate result compared with other estimators.
The Landy-Szalay estimator~$\xi_{\rm obs}(\theta)$ is given as a function of a
separation angular scale $\theta$ through
\begin{equation}
	\xi_{\rm obs}(\theta) 
   = 
   \frac{%
   n^{D}_{\rm clu}I^{D}_{\gamma} - 
   n^{D}_{\rm clu}I^{R}_{\gamma} - 
   n^{R}_{\rm clu}I^{D}_{\gamma} + 
   n^{R}_{\rm clu}I^{R}_{\gamma}}%
   {n^{R}_{\rm clu}I^{R}_{\gamma}}, 
\end{equation}
where $n_{\rm clu}=n_{\rm clu}(\bs{\theta}_1)$ is the number of clusters 
at a cluster position, $\bs{\theta}_1$. 
$I_{\gamma}=I_{\gamma}(\bs{\theta}_2)$ is the number intensity of $\gamma$-ray 
at $\bs{\theta}_2$ where $|\bs{\theta}_1-\bs{\theta}_2|=\theta$.
The superscripts $D$ and $R$ represent the actual data 
(i.e. the CAMIRA cluster number count and the UGRB intensity)
and the random data. 
We use the random CAMIRA catalogue containing 185,459 clusters as random data of the CAMIRA cluster number count. 
The random CAMIRA catalogue is made by running the CAMIRA for 
randomly distributed mock samples of galaxies within the same fields as data.   
For the random map of the UGRB, we generate the random intensity map assuming 
photon obeys a Poisson distribution with total number of photons being 
one hundred times larger than what observed. 
\subsection{covariance}
\label{ssec:covariance}
%
To evaluate the error on the cross-correlation estimator,
we use the jackknife method~\citep{Scranton-Johnston:2002}.
Here we divide both CAMIRA catalogue and the UGRB map into 21~subregions.
Individual subregion spans 
$3.5\times 3.5$[deg$^2$] area 
on average, which is sufficiently larger than the scale of interest.
The covariance matrix is then given by
\begin{equation}
\label{eq:covariance}
{\bf C}_{ij} = \frac{M-1}{M} 
\sum^{M}_{k=1}
\left[\widehat{\xi_{k}(\theta_i)} - \overline{\xi(\theta_i)} \right] \times 
\left[\widehat{\xi_{k}(\theta_j)} - \overline{\xi(\theta_j)} \right], 
\end{equation}
where $M$ is the number of the jackknife subsamples, $M=21$ in our case,
and $\widehat{\xi_{k}(\theta)}$ is the correlation function 
for the $k$-th jackknife re-sample
where we remove the $k$-th subregion from the entire region.
Mean correlation function is obtained by averaging over all different jackknife 
re-samples,
\begin{equation}
	\overline{\xi(\theta)}
    = 
    \frac{1}{M}\sum_k^M \widehat{\xi_k(\theta)}
\end{equation}

In Figure~\ref{fig:ccf_pro}, we show the profiles of the
cross-correlations for clusters in all redshifts, 
low redshifts~($z<0.6$) and high redshifts~($z>0.6$). 
One can clearly see the cross-correlation signals within $\sim 1$~degree, for the all-redshift clusters and the low-redshift cases. The signals for low-redshift clusters extend wider than those for all clusters. The excess at the cluster position is independent on the Galactic foreground model. Contrarily, the cross-correlation signals for high-redshift clusters are rather weak. In particular, with the baseline model, it seems that the signal is consistent with zero. The difference among the different foreground models is within the one-sigma error bars. Therefore, with the current data sets of the high-redshift clusters, we cannot conclude the detection or non-detection of the cross-correlation signal and cannot make a further discussions on the dependence of the signal on the Galactic foreground models. Note that as we will describe in \ref{ssec:detection}, there are strong correlations among different angular bins.
Also note that the error estimation of the jackknife method sometimes brings misestimation but it is quite hard to perform full simulation of $\gamma$-ray diffuse sky as we do not understand the origin of the diffuse emission. Those treatments are beyond the scope of this paper and we emphasise that the results are all based on the error estimation from the jackknife method.

\subsection{Detection}
\label{ssec:detection}
\begin{figure*}
	\begin{tabular}{cc}
		\includegraphics[width=0.5\linewidth]{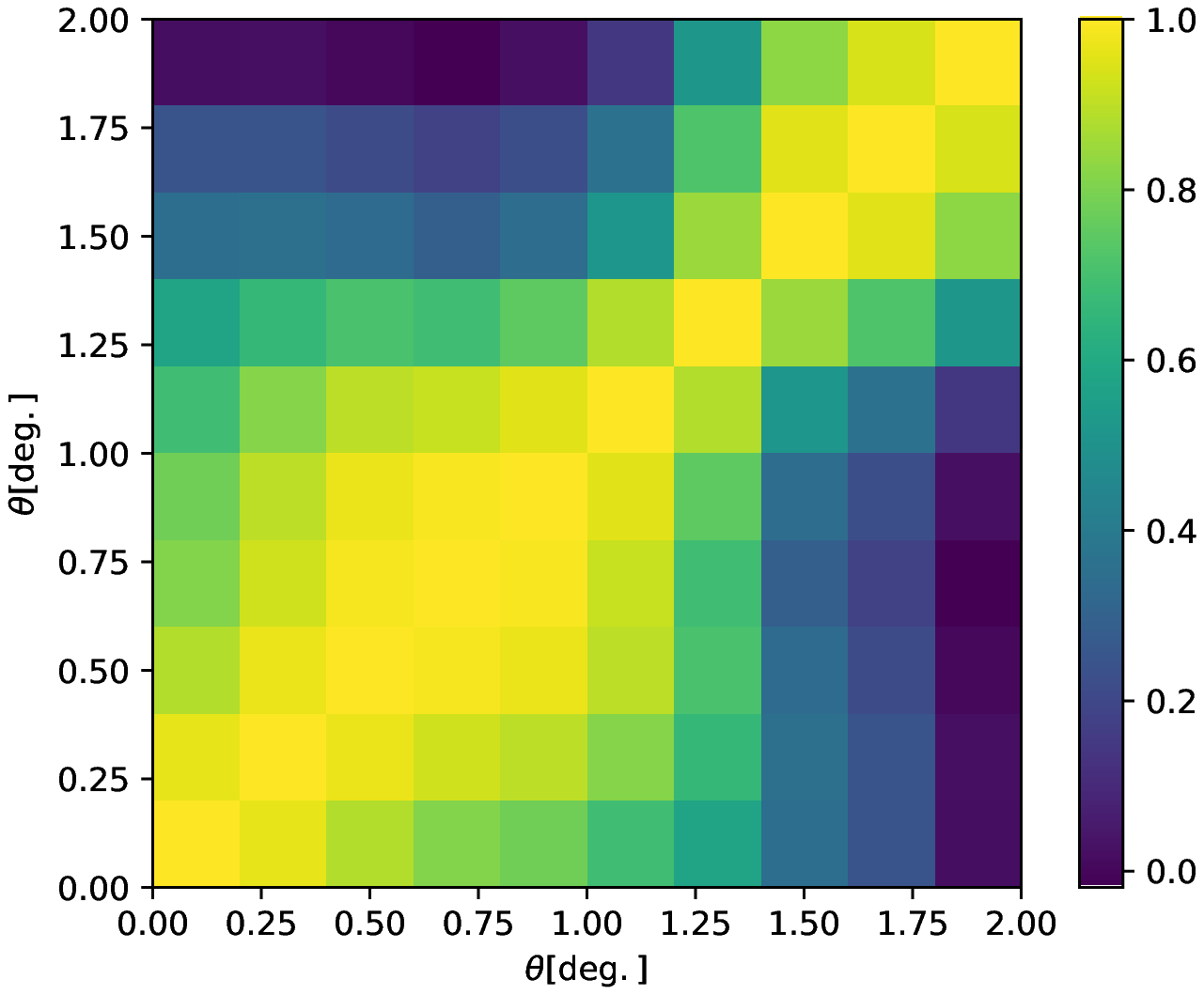}&
		\includegraphics[width=0.5\linewidth]{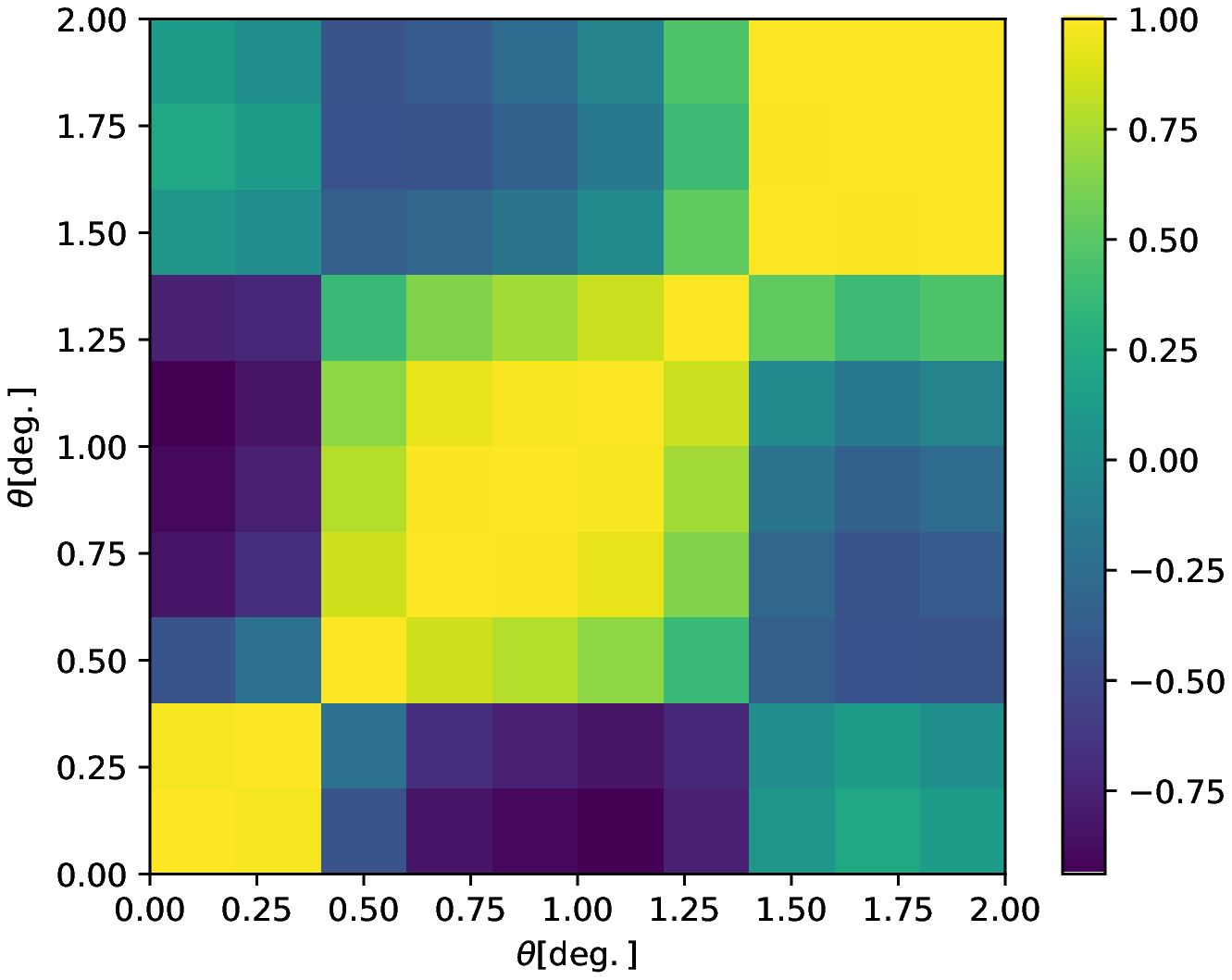}\\
    \end{tabular}
    \caption{(\textit{Left}) Covariance matrix of cross correlation for the base-line foreground model, normalised by diagonal terms, ${\rm C}_{ij}/\sqrt{{\rm C}_{ii} {\rm C}_{jj}}$. Off diagonal terms are not negligible which mean a significant correlation among different angular scales. (\textit{Right}) Pseudo-inverse of the covariance matrix remaining largest four eigenvalues (see text for details), normalised again by diagonal terms.
    \label{fig:cov}}
\end{figure*}

%
Here we estimate the statistical significance of the signal for all-redshift, low-redshift and high-redshift galaxy cluster sets. To evaluate the significance we adopt the standard $\chi^{2}$ statistics with the null hypothesis taking into account a possible correlation among different angular bins, 
\begin{equation}
	\chi^{2}_{\rm mod} = \sum_{i,j} 
    \left[ \overline{\xi(\theta_i)} - \xi_{\rm mod}(\theta_i) \right] 
    \widetilde{\bf C}^{-1}_{ij} 
    \left[ \overline{\xi(\theta_j)} - \xi_{\rm mod}(\theta_j)\right],
\label{eq:chi_mod}
\end{equation}
where $\xi_{\rm mod}(\theta)$ is a hypothetical prediction for the cross-correlation function, which meanwhile we assume to be zero to find the detection significance and is later given in Section~\ref{sec:implication} to compare the signal with the theoretically predicted model.
The matrix $\widetilde{\bf C}^{-1}$ is a pseudo-inverse of the covariance matrix. As can be seen in Figure~\ref{fig:cov}, there are significant correlations between different angular scales. In general, if there is a strong correlation,  the matrix inversion is noisy. To regularize the matrix, we apply a singular value decomposition (SVD) to obtain the pseudo-inverse of the covariance matrix, which is numerically stable. The covariance matrix can be decomposed as
\begin{equation}
	{\bf C} = {\bf u} ~ \bs{\Sigma} ~ {\bf v}^{\rm T},
\end{equation}
where $\Sigma$ is a diagonal matrix which consists of eigenvalues of the covariance matrix $\lambda_i$ arranged in a descending order. We then remove the eigenvalues small enough compared to the largest eigenvalue, which makes inversion noisy. By looking at the numerical convergence, we decide largest four eigenvalues are meaningful and others are not informative. Then the pseudo inverse matrix is given as
\begin{equation}
	\widetilde{\bf C}^{-1}
   =
   {\bf v} ~ \bs{\Sigma}^{\prime -1} ~ {\bf u}^{\rm T},
\end{equation}
where
\begin{equation}
	\bs{\Sigma}^{\prime -1} 
    =
   {\rm diag}
   \left(
   	\frac{1}{\lambda}_1,
   	\cdots,
   	\frac{1}{\lambda}_k,
    0,
    \cdots,
    0
   \right),
\end{equation}
and $\lambda_i$ are the eigenvalues of covariance matrix satisfying $\lambda_{i} > \lambda_{i+1}$ for all $i$. The smallest eigenvalue which is not negligible is $\lambda_k$.

The detection significance which is summarised in Table \ref{tab:ccf_sig} can be obtained by setting $ \xi_{\rm mod}(\theta_i)=0$  in Eq~(\ref{eq:chi_mod}), 
assuming the probability function obeys Gaussian distribution. 
We find that the cross-correlation signal is around $2\sigma$ or smaller, which means the marginal evidence for the $\gamma$-ray association to the cluster of galaxies. In addition, the significance fluctuates depending on the foreground model we assumed and thus the detection significance is rather weak.

We note that although it is empirically known that for $10\times 10$ covariance, it is required to have 100 jackknife subregions.
However, in our dataset, due to the limited survey area of HSC and the scale of our interest which extends up to 2 degrees, it is difficult to patch the sky into 100 subregions. Therefore, we revisit the analysis to derive the significance with coarser binning of $4\times 4$ covariance, which can be well measured by 21 jackknife sampling. We find that the obtained significance is different only by less than 10 per cent from those for 10 binning, which implies that our measurement of $10\times 10$ covariance with 21 subsampling would be reasonable.

As we described in section \ref{ssec:hsc_photo}, we have used the cluster catalogue aggressively, $\hat{N}_{\rm mem}>10$; however, it may contain some fake clusters due to the projection effect. In order to see the robustness of our analysis, we repeat our analysis for the sample $\hat{N}_{\rm mem}>15$. For the sample $\hat{N}_{\rm mem}>15$, we find smaller amplitude of cross correlation at low-z and larger at high-z; however, they are still consistent with each other within 1$\sigma$ error bars.
The subtle mass and redshift dependencies of the signal imply that the CAMIRA clusters likely contaminated by the fake clusters at high-redshift especially at less massive end so that our signal for $\hat{N}_{\rm mem}>10$ are more contaminated by the fake clusters, while it is less affected by the fakes at low-redshift.

\iffigure
\begin{figure}
  \centering
    \begin{tabular}{c}
      \includegraphics[width=0.65\linewidth, angle=-90]{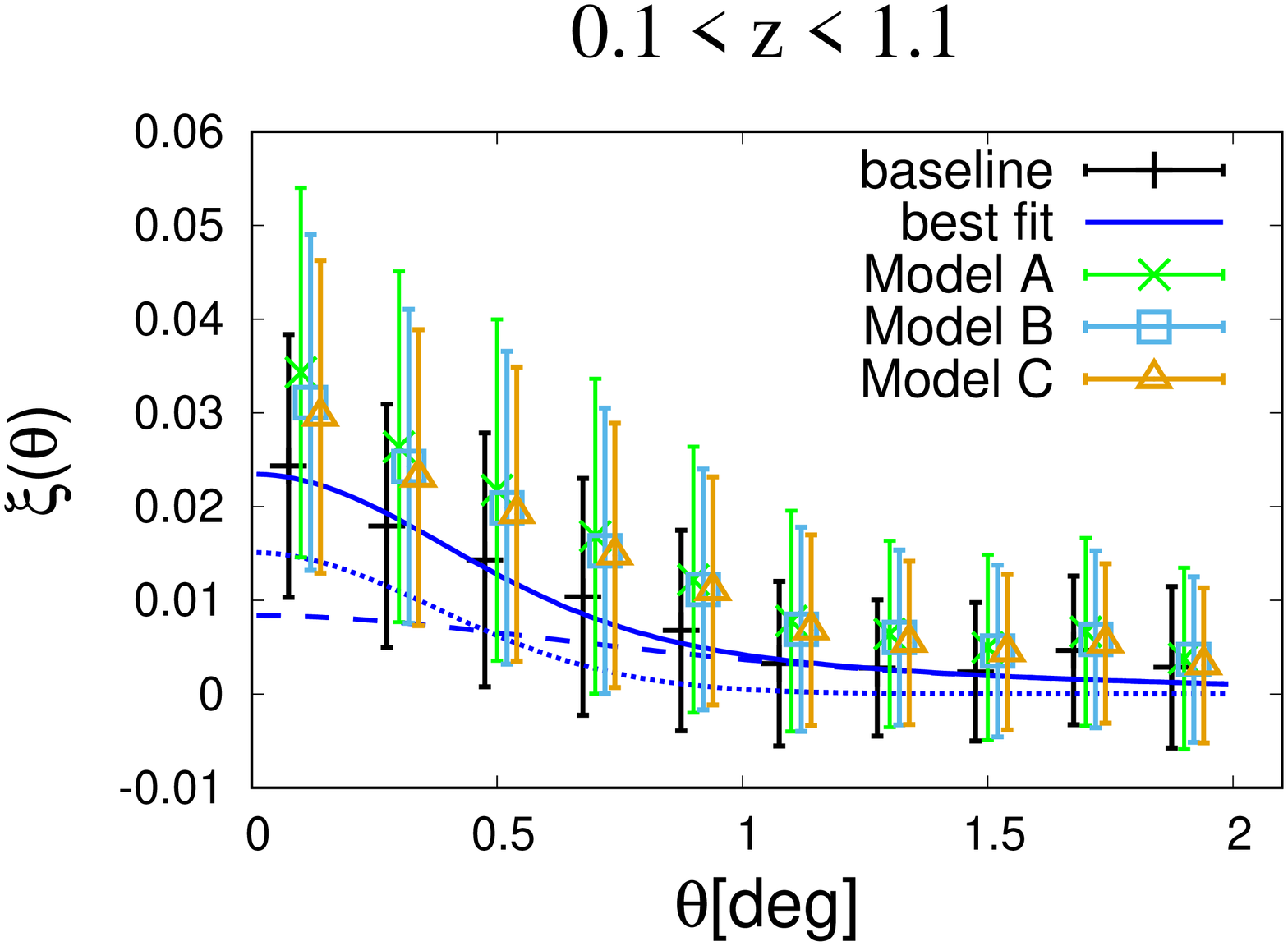} \\
      \includegraphics[width=0.65\linewidth, angle=-90]{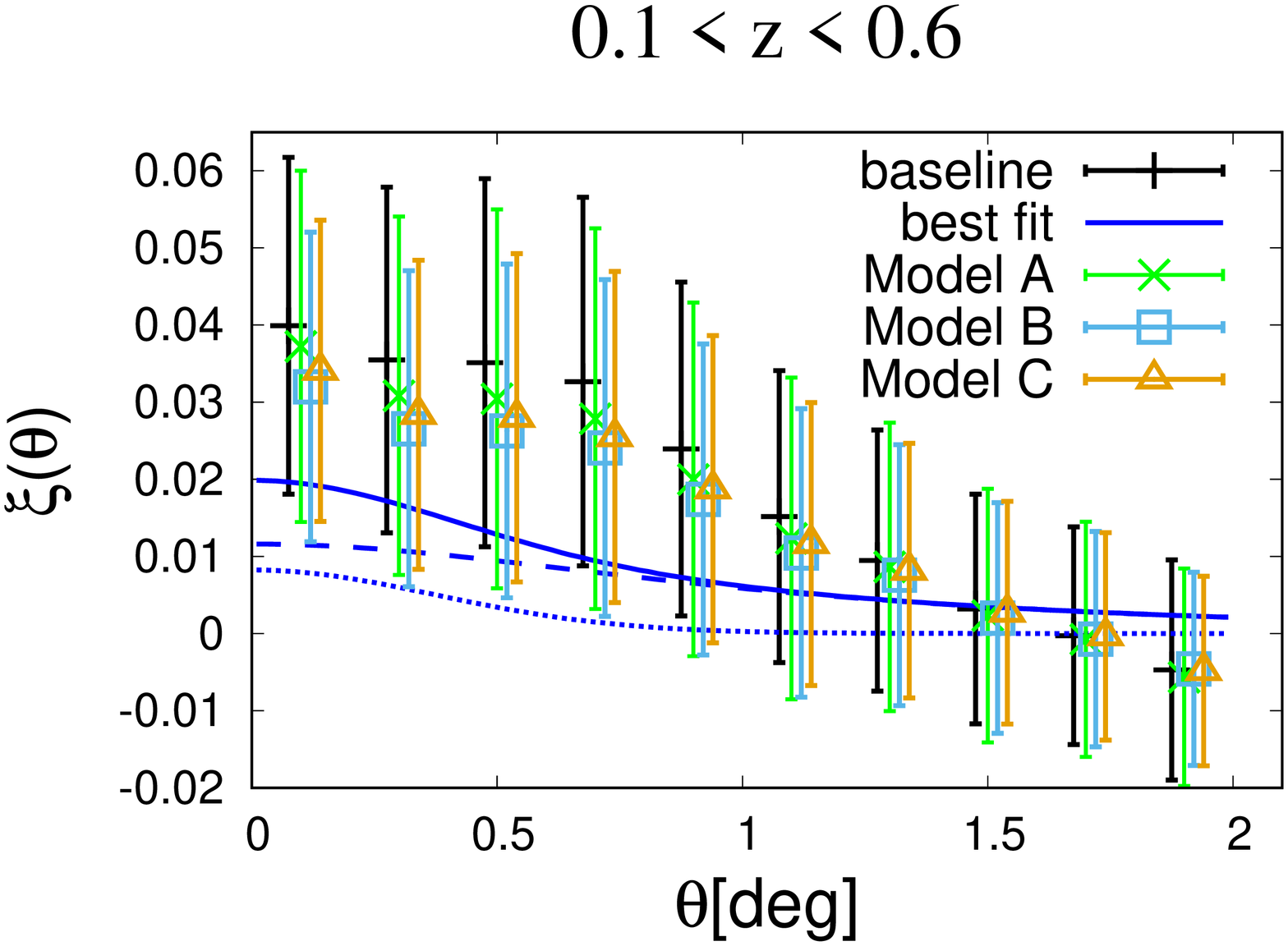} \\
      \includegraphics[width=0.65\linewidth, angle=-90]{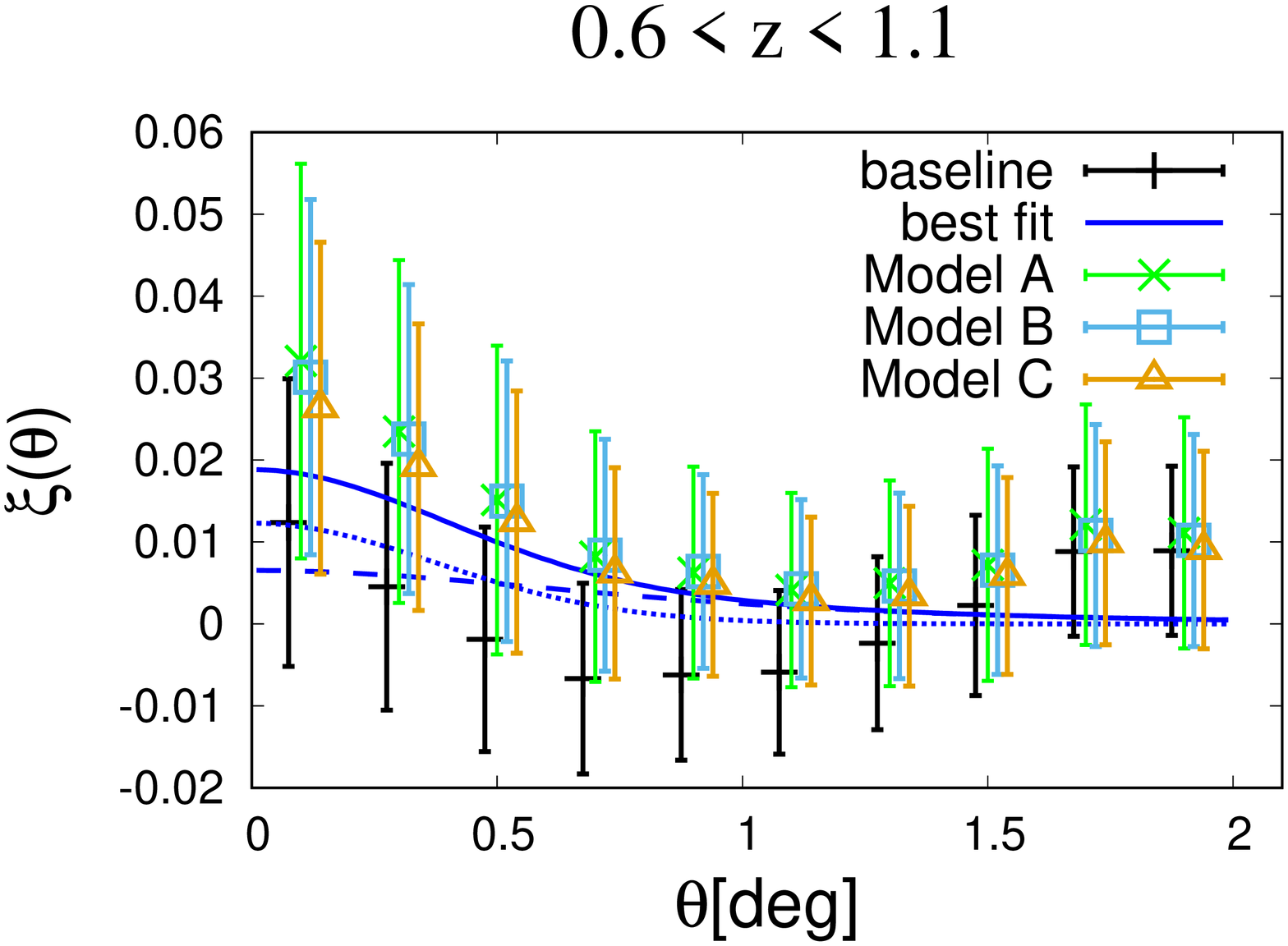} \\
    \end{tabular}
    \caption{
    Cross correlation signal of $\gamma$-ray map and cluster distribution for all-(\textit{top}) , low redshift-(\textit{middle}) and high redshift-(\textit{bottom}) clusters, respectively. Symbols with error bars are measured angular cross correlation with different symbols correspond to different foreground models. Solid lines are best-fitted theoretical predictions to the data corrected with the baseline foreground model. The theoretical model can be decomposed into to components: astronomical objects such as blazars, star-forming galaxies and radio galaxies (\textit{dashed}) and shot noise term (\textit{dotted}).
For more details about the theoretical prediction, refer to the text in section~\ref{sec:implication}.
\label{fig:ccf_pro}}
\end{figure}
\fi

\begin{table}
  \centering
    \begin{tabular}{ccccc} \hline\hline
    redshift range & baseline & Model~A & Model~B & Model~C \\ \hline
     $0.1~<~z~<~1.1$ & 2.2 & 2.0 & 2.0 & 2.0 \\
     $0.1~<~z~<~0.6$ & 2.2 & 2.1 & 2.1 & 2.3 \\ 
     $0.6~<~z~<~1.1$ & 1.9 & 1.6 & 1.6 & 1.6 \\ \hline \hline
    \end{tabular}
  \caption{
     Statistical significances for detecting the cross correlation 
     with different Galactic foreground models and redshift ranges of clusters.
     Note that we use 1942 low-redshift clusters and 2519 high-redshift clusters. 
     }
  \label{tab:ccf_sig}
\end{table}


\section{Implication}
\label{sec:implication}
The UGRB is expected to be cumulative $\gamma$-ray 
emissions from unresolved extragalactic sources as well as some diffuse processes.
Blazars are one of the most abundant point sources in GeV $\gamma$-ray energy and 
known to be a main contributor to extragalactic $\gamma$-ray emissions.
The latest model of $\gamma$-ray spectrum and luminosity function of blazars 
can naturally explain about 70 per~cent of the extragalactic $\gamma$-ray intensity above 1 GeV \citep[e.g.][]{2015ApJ...800L..27A}.
Other possible candidates contributing to UGRB include star forming galaxies \citep[e.g.][]{2007ApJ...654..219T}
and radio galaxies \citep[e.g.][]{2011ApJ...733...66I}.
Here, we consider a simple model of cross correlation assuming
that blazars, star-forming galaxies, and radio galaxies are biased tracers of large-scale structures in the Universe.

The UGRB intensity along a given direction $\bvec{\theta}$ can be written as
\beqa
I_{\gamma}(\bvec{\theta}) = \sum_{X} \int {\rm d}\chi\, W_{\gamma, X}(\chi)\, g_{X}(r(\chi)\bvec{\theta}, \chi),
\label{eq:UGRB_intensity}
\eeqa
where 
$\chi(z)$ 
denotes the radial comoving distance (function of redshift $z$), 
$r(\chi)$ is the comoving angular diameter distance,
$W_{\gamma, X}(\chi)$ is the window function for population $X$,
and $g_{X}(\bvec{x})$ is the relevant density field of the $\gamma$-ray source $X$.
We here define $W_{\gamma, X}(\chi)$ so that the mean $\gamma$-ray intensity can be expressed as
$\sum_{X}\, \int {\rm d}\chi\, W_{\gamma, X}(\chi)$.
We compute the window function $W_{\gamma, X}$ as follows in \citet{2018arXiv180210257S}.
The model includes $\gamma$-ray luminosity function and energy spectrum for different source populations 
and $\gamma$-ray attenuation during propagation owing to pair creation on diffuse extragalactic photons \citep{2012MNRAS.422.3189G}.
For blazars, we consider a parametric description of $\gamma$-ray luminosity function and  
energy spectrum as developed in \citet{2015ApJ...800L..27A}.
For star-forming galaxies, we assume a correlation of luminosity between $\gamma$ ray and infrared
to derive their $\gamma$-ray luminosity function \citep{2012ApJ...755..164A}. The energy spectrum of star-forming galaxies is set to be $E^{-2.2}$ \citep{2012ApJ...755..164A}.
For radio galaxies, we follow the model in \citet{2014ApJ...780..161D}, which has established a correlation
between the $\gamma$-ray luminosity and the radio-core luminosity at 5 GHz. 
We assume an average spectral index of 2.37 for radio galaxies.
To define the ``unresolved" components for population $X$, 
we set the $\gamma$-ray flux limit of $2\times10^{-9}\, {\rm cm}^{-2}\, {\rm s}^{-1}$ above 100 MeV.

The angular number density field of CAMIRA clusters can be expressed as
\begin{eqnarray}
&&n_{\rm clu}(\bvec{\theta}) = \int {\rm d}\chi\, \chi^2 \int {\rm d}M\, S\left(M, z\left(\chi\right)\right)\, \nonumber \\
&&\hspace{9em}
\times \bar{n}\left(M, z\left(\chi\right)\right)
\left[1+\delta_h(\chi \bvec{\theta}, \chi)\right],
\label{eq:n_CAMIRA}
\end{eqnarray}
where $\bar{n}(M,z)$ represents the halo mass function, $S(M,z)$ is the selection function for CAMIRA clusters,
and $\delta_h(\bvec{x})$ is the overdensity field of three-dimensional halo number density.
As in Section \ref{ssec:hsc_photo}, we assume $S=1$ for $M\ge10^{13.5}\, h^{-1}\, M_{\odot}$ and $S=0$ otherwise in this paper. 
We adopt the model of $\bar{n}$ as in \citet{2008ApJ...688..709T} 
for the spherical overdensity parameter $\Delta=200$ with respect to mean cosmic matter density.
Under a flat-sky approximation, we can express the cross correlation between $n_{\rm clu}$ and $I_{\gamma}$ in Fourier space as
\citep[also see][]{Branchini+:2017},
\beqa
P_{{\rm c}\gamma}(\ell) = \sum_{X} \int \frac{{\rm d}\chi}{r(\chi)^2}\, W_{\gamma, X}(\chi) \, W_{\rm clu}(\chi) \,
P^{(\rm 3D)}_{{\rm h}X}\left(\frac{\ell}{r(\chi)}, z(\chi)\right), \label{eq:cross_Pell}
\eeqa
where $\ell$ is the multipole and $P^{(\rm 3D)}_{{\rm h}X}$ represents the three-dimensional cross power spectrum
between $g_{\gamma, X}$ and $\delta_{h}$ (see Eqs.~(\ref{eq:UGRB_intensity}) and (\ref{eq:n_CAMIRA})).
In Eq.~(\ref{eq:cross_Pell}), we define the effective window function for CAMIRA clusters as
\beqa
W_{\rm clu}(\chi) \equiv \chi^2 \int {\rm d}M\, S\left(M, z\left(\chi\right)\right)\, 
\bar{n}\left(M, z\left(\chi\right)\right).
\eeqa
On degree scales, linear approximation of the evolution in two fields $g_{\gamma, X}$ and $\delta_{h}$
will be valid. In contrast, we expect the shot noise term arising from the finite sampling of 
CAMIRA clusters and unresolved $\gamma$-ray sources for small scales.
In this case, the three-dimensional power spectrum can be approximated as
$P^{(\rm 3D)}_{{\rm h}X}(k, z)= \bar{b}_{\rm h}(z) \bar{b}_{{\rm eff},X}(z)  P_{\rm L}(k, z) + {\cal A}$,
where $P_{\rm L}(k, z)$ represents the linear matter power spectrum at redshift $z$,
$\bar{b}_{\rm h}$ is the linear halo bias including the selection effect in halo mass, 
$\bar{b}_{{\rm eff}, X}$ is the effective bias factor of $\gamma$-ray population $X$,
and ${\cal A}$ is the shot noise term.
In this paper, we compute $\bar{b}_{\rm h}$ as
\beqa
\bar{b}_{\rm h}(z) = \frac{\int {\rm d}M\, S(M,z)\, \bar{n}(M,z) \, b(M,z)}{\int {\rm d}M\, S(M,z)\, \bar{n}(M,z)},
\eeqa
where $b$ is the linear bias and we adopt the model of $b$ in \citet{2010ApJ...724..878T}.
Unfortunately $\bar{b}_{{\rm eff}, X}$ is still uncertain and 
it should be constrained by clustering analyses of gamma rays with large-scale structures in practice.
For simplicity, we adopt the fiducial model of $\bar{b}_{{\rm eff}, X}$ in \citet{2018arXiv180210257S}.
We leave possible constraints of $\bar{b}_{{\rm eff}, X}$ by our measurements for our future work.
It is also worth mentioning that there exist other possible sources to generate the cross correlation.
They include the $\gamma$-ray 
emissions from intra-cluster medium \citep[e.g.][]{1998APh.....9..227C}
and dark matter annihilations \citep[e.g.][]{1996PhR...267..195J}.
Our measurement can be useful to constrain these possible $\gamma$-ray emissions and we plan to explore them in the future.

Our observable $\xi$ is related with $P_{{\rm c}\gamma}(\ell)$
through Fourier transform: 
\beqa
\xi(\theta) = \frac{1}{\langle n_{\rm clu}\rangle\langle I_{\gamma}\rangle}\int \frac{{\rm d}^{2} \ell}{(2\pi)^2}\, 
\exp\left[i\bvec{\ell} \cdot \bvec{\theta}\right]\, P_{{\rm c}\gamma}(\ell)
\hat{W}(\ell,\theta_{G}),
\eeqa
where $\langle\cdots\rangle$ represents the average over a sky and 
$\hat{W}$ is the Fourier counterpart of two-dimensional Gaussian smoothing filter with the smoothing scale of $\theta_G$.
Although Eq.~(\ref{eq:cross_Pell}) does not include the smearing effect by $\gamma$-ray point spread function (PSF),
we include the effect by convolving the $W_{\gamma, X}$ with the $\gamma$-ray PSF 
when comparing our prediction with observation \citep[see][for details]{2014PhRvD..90f3502S}.
In summary, our model of $\xi$ has single free parameter $\cal A$, while we fix the cross correlation coming from large-scale
clustering between CAMIRA clusters and $\gamma$-ray sources.

Blue solid lines in figure \ref{fig:ccf_pro} show 
the best-fit model of cross correlation.
We define the best-fit model to minimise the chi-squared value as in Eq.~(\ref{eq:chi_mod}) by varying the parameter. 
We find our simple model with single free parameter 
can reasonably explain the observed correlation regardless of selections in cluster redshift.
Systematic effects due to imperfect modeling of Galactic $\gamma$-ray foreground are found to be unimportant for the current analysis.
The minimum chi-squared values and the number intensity of the shot noise term 
giving the minimum $\chi^2_{\rm mod}$ in our analysis are summarised in 
Table~\ref{tab:ccf_chi} and Table \ref{tab:bestfit_A}, respectively. 
In the case of low-redshift clusters, astronomical sources contribute 
largely for the correlation and the shot noise term has a weaker effect. 
In contrast, in the case of all clusters or high-redshift clusters, 
the shot noise term is dominant for model prediction on small scale~($<\sim 0.5^{\circ}$). 
We note that the best-fitting amplitudes of the shot noise term are larger than the  ones in \citet{Branchini+:2017} by a factor of 10
;however, due to the large uncertainties, it is still consistent within $1\sigma$ errors.
We also note that the theoretical model includes uncertainties for 
$\gamma$-ray sources and a bias parameter $b_{\rm eff}$.

Due to the small number of samples for the $\gamma$-ray sources, it is yet difficult to discuss the effect of flux uncertainties on the model prediction but according to the current limited samples, the uncertainties on the $\gamma$-ray sources ranges from 0.2 to 2
\citep{2015ApJ...800L..27A, 2012ApJ...755..164A, 2011ApJ...733...66I}.
The model for the bias also includes large uncertainties; we have no reliable Halo Occupation Distribution model for blazars yet. \cite{2014ApJ...797...96A} measured angular correlation of BL Lacs and flat-spectrum radio quasars and found that the linear biases for them are $1.84\pm0.25$ and $3.30\pm 0.41$, respectively. From this study, we can infer  the magnitude of the bias uncertainty is about 50 per cent.

Since the angular resolution in \emph{Fermi} UGRB is of an order of degree,
the effective number of degrees of freedom can be smaller than figure \ref{fig:ccf_pro} actually shows. 
To investigate the nature of UGRB with our cross-correlation function in details,
the cross correlation measurement in Fourier space will be essential and we will work on Fourier-space analysis in the future.

\begin{table}
  \begin{center}
    \begin{tabular}{lcccc} \hline\hline
    redshift range & Baseline & Model~A & Model~B & Model~C \\ \hline
     $0.1~<z<~1.1$ & 0.30 & 0.29 & 0.28 & 0.27 \\
     $0.1~<z<~0.6$ & 2.2 & 1.9 & 1.8 & 2.2 \\
     $0.6~<z<~1.1$ & 2.5 & 0.79 & 0.76 & 0.78 \\ \hline \hline
    \end{tabular}
  \end{center}
  \caption{
     The minimum chi-squared values in our comparison 
     with model and observed correlation $\chi^{2}_{\rm mod}$.
     We summarise the results for three different cluster redshift ranges
     for different Galactic foreground models.
     Note that effective number of degree-of-freedom is 3 for all cases.
     }
  \label{tab:ccf_chi}
\end{table}

\begin{table}
  \begin{center}
    \begin{tabular}{lcccc} \hline\hline
    redshift range & Baseline & Model~A & Model~B & Model~C \\ \hline
     $0.1~<z<~1.1$ & 5.1 & 6.3 & 5.6 & 5.2 \\
     $0.1~<z<~0.6$ & 2.8 & 4.4 & 3.6 & 3.6 \\
     $0.6~<z<~1.1$ & 4.2 & 6.9 & 6.3 & 5.6 \\ \hline \hline
    \end{tabular}
  \end{center}
  \caption{
     The best fit value for the amplitude of the shot noise term~($10^{-9}{\rm cm^{-2}s^{-1}sr^{-1}}$). 
     }
  \label{tab:bestfit_A}
\end{table}


%
\section{Summary}
\label{sec:conclusion} 
In this paper, we have investigated the cross-correlation signal
between the UGRB,~the PASS8 \emph{Fermi}-LAT data, and
the HSC galaxy clusters,~the CAMIRA cluster catalogue.
To evaluate the cross-correlation signals,
we have performed both the stacking analysis and the cross-correlation analysis.
We have evaluated the
cross-correlation signal quantitatively by the cross-correlation analysis,
while the stacking analysis is used to confirm
the validity of the cross-correlation analysis.

In both analyses,
we have found evidence of the cross-correlation signal.
The signal appears as a few~per~cent excess of the
photon number intensity to the background.
Although we have adopted four models of the
Galactic diffuse emission
to subtract the foreground contamination,
the signal does not depend on the foreground model.
The statistical significance of the signal detection 
reaches 2.0-2.3$\sigma$ significance independently of any of our Galactic foreground models.

The cross-correlation analysis has shown that
the signal extends up to $\sim 1^{\circ}$ from cluster centres.
This angular scale corresponds to $5 {\rm Mpc}/h$ (at $z=0.1$) to 
$40 {\rm Mpc}/h$ (at $z=1.1$) which is well
beyond the typical scale of cluster.
The size of PSF for the \emph{Fermi}-LAT data varies from $0.1^{\circ}$ to 
$1^{\circ}$ for 100 GeV to 1 GeV and we find that our cross correlation 
signal is fully dominated by the lower energy photons (1-5 GeV).
Therefore, we conclude that while the extension of signal is mainly dominated by the
PSF of the $\gamma$-ray photons, we may not exclude the possibility 
that this signal extension is due to the over-densities around cluster regions
\citep[e.g.][]{Branchini+:2017}.

We have also studied the redshift dependence of the cross-correlation
signal, by dividing the cluster catalogue into two redshift bins,
low-redshift clusters~$(0.1<z<0.6)$ and high-redshift clusters~$(0.6<z<1.1)$.
We have found that the cross-correlation signal with low-redshift sample 
is stronger than that with the high-redshift sample.
The detection significances are $2.1-2.3 \sigma$, $1.6-1.9 \sigma$, and $2.0-2.2 \sigma$ 
for low-redshift, high-redshift, and all-redshift samples, respectively.
With our current dataset, we do not claim a significant detection of the cross correlation between $\gamma$-ray sources and cluster of galaxies, given that the statistical significance fluctuates depending on the foreground model we assumed.

We see that the signals for only $\hat{N}_{\rm mem}>15$ are still consistent with all samples of $\hat{N}_{\rm mem}>10$ and the obtained significance is not largely affected by this choice of the minimum mass of the clusters.

To study the source of the cross-correlation signal,
we have compared the measured cross-correlation function
with a theoretical model.
The model includes the contributions
from blazars, star-forming galaxies and radio galaxies as $\gamma$-ray
emitters in a cluster.
It seems that the detected signal is fairly consistent with
the theoretically predicted model.
We leave a more detailed modeling of the astronomical origins of the 
$\gamma$-ray emitters as well as the exploration to the light from 
dark matter for future work.

In this analysis, we used $\sim 4000$ clusters from the current HSC 
CAMIRA cluster catalogue
and $\sim 200~{\rm deg}^{2}$~UGRB map~(the number of $\gamma$-ray photons is $\sim 3000$).
We repeat the same analysis for half of our survey region and find that the error scales by a factor of $\sqrt{2}$, which means the error is fully explained by the cosmic variance.
As more data is accumulated, our detected signal can
become more robust with smaller error bars in the future.
For example, the number of CAMIRA clusters is expected to increase  as the HSC observation area will increase by at least three times,  the statistical error can be reduced by a factor of~$1/\sqrt{3}$.
If the same signal appears in future data,
the significances of the such correlation signals
will be improved to $3\sigma$ levels even for high-redshift clusters. 
The accumulation of the data in both galaxy clusters and $\gamma$-ray
photons will allow us to perform further statistical analysis
including the one with multiple redshifts or energy binnings.
Such analyses may not only reveal more details
of the cross-correlation of the UGRB with clusters
but also constrain the nature of dark matter as a
source of $\gamma$-rays through their annihilations, decays or 
radiation as primordial black-holes evaporation. 

\section*{acknowledgements}

We would like to thank Naoshi Sugiyama, Kenji Hasegawa and Teppei Minoda for useful discussions and comments for this work and also
to Chien-Hsiu Lee for careful reading of
our manuscript and giving us useful comments.
AN is supported in part by MEXT KAKENHI Grant Number 16H01096. 
SH is supported by the U.S.~Department of Energy under award number DE-SC0018327. 
HT is supported by JSPS KAKENHI Grant Number 15K17646, 17H01110. 

The Hyper Suprime-Cam (HSC) collaboration includes the astronomical
communities of Japan and Taiwan, and Princeton University.
The HSC instrumentation and software were developed by the National
Astronomical Observatory of Japan (NAOJ), the Kavli Institute for the
Physics and Mathematics of the Universe (Kavli IPMU), the University
of Tokyo, the High Energy Accelerator Research Organization (KEK), the
Academia Sinica Institute for Astronomy and Astrophysics in Taiwan
(ASIAA), and Princeton University.  Funding was contributed by the FIRST 
program from Japanese Cabinet Office, the Ministry of Education, Culture, 
Sports, Science and Technology (MEXT), the Japan Society for the 
Promotion of Science (JSPS),  Japan Science and Technology Agency 
(JST),  the Toray Science  Foundation, NAOJ, Kavli IPMU, KEK, ASIAA,  
and Princeton University.

The Pan-STARRS1 Surveys (PS1) have been made possible through
contributions of the Institute for Astronomy, the University of
Hawaii, the Pan-STARRS Project Office, the Max-Planck Society and its
participating institutes, the Max Planck Institute for Astronomy,
Heidelberg and the Max Planck Institute for Extraterrestrial Physics,
Garching, The Johns Hopkins University, Durham University, the
University of Edinburgh, Queen's University Belfast, the
Harvard-Smithsonian Center for Astrophysics, the Las Cumbres
Observatory Global Telescope Network Incorporated, the National
Central University of Taiwan, the Space Telescope Science Institute,
the National Aeronautics and Space Administration under Grant
No. NNX08AR22G issued through the Planetary Science Division of the
NASA Science Mission Directorate, the National Science Foundation
under Grant No. AST-1238877, the University of Maryland, and Eotvos
Lorand University (ELTE).

This paper makes use of software developed for the Large Synoptic
Survey Telescope. We thank the LSST Project for making their code
available as free software at http://dm.lsst.org.

\bibliographystyle{mn2e}
\bibliography{bibdata}

\end{document}